\documentclass[11pt,aps,prd,nofootinbib,tightenlines,preprintnumbers,amsmath,amssymb,longbibliography,notitlepage]{revtex4-1}
\usepackage{epsfig}
\usepackage{graphicx}
\usepackage{latexsym}
\usepackage{booktabs}
\usepackage{graphicx}% Include figure files
\usepackage{color}
\usepackage{footnote}
\usepackage{hyperref}

\definecolor{darkred}{rgb}{0.55, 0.0, 0.0}
\definecolor{crimsonglory}{rgb}{0.75, 0.0, 0.2}

\newcommand{\be}{\begin{equation}}
\newcommand{\ee}{\end{equation}}
\newcommand{\ba}{\begin{eqnarray}}
\newcommand{\ea}{\end{eqnarray}}
\newcommand{\bi}{\begin{itemize}}
\newcommand{\ei}{\end{itemize}}

\newcommand{\<}{\langle}
\renewcommand{\>}{\rangle}
\newcommand{\eq}{Eq.~}

\newcommand{\fig}{Fig.~}

\newcommand{\la}{\label}

\begin{document}
\title{Charge transport and vector meson dissociation \\ across the thermal phase transition in lattice QCD 
with two light quark flavors}
% { in light two-flavor lattice QCD}
% \title{Thermal dilepton rates, electrical conductivity and vector meson dissociation
%  across the deconfinement phase transition in light two-flavor lattice QCD}

\author{Bastian B. Brandt$^1$, Anthony Francis$^2$, Benjamin J\"ager$^3$ and Harvey B. Meyer$^4$\vspace{0.2cm}}

\affiliation{
$^1$Institut f\"ur theoretische Physik, Universit\"at Regensburg, D-93040 Regensburg, Germany\\
$^2$Department of Physics \& Astronomy, York University, 4700 Keele St, Toronto, ON M3J 1P3, Canada\\
$^3$Department of Physics, College of Science, Swansea University, SA2 8PP Swansea, United Kingdom\\
$^4$PRISMA Cluster of Excellence, Institut f\"ur Kernphysik and Helmholtz~Institut~Mainz,
Johannes Gutenberg-Universit\"at Mainz, D-55099 Mainz, Germany 
\vspace{0.2cm}
%        E-mail: \email{francis@kph.uni-mainz.de},
}

\preprint{MITP/15-019}

\date{\today}
% \pacs{12.38.Gc, 12.38.Mh, 25.75.-q}

%%%%%%%%%%%%%%%%%%%%%%%%%%%%%%%%%%%%%%%%%%%%%%%%%%%%%%%%%%%%%%%%
%%%%%%%%%%%%%%%%%%%%%%%%%%%%%%%%%%%%%%%%%%%%%%%%%%%%%%%%%%%%%%%%
\begin{abstract}
We compute and analyze correlation functions in the isovector vector
 channel at vani\-shing spatial momentum across the
deconfinement phase transition in lattice QCD. The simulations are
carried out at temperatures $T/T_c=0.156, 0.8, 1.0, 1.25$ and $1.67$
with $T_c\simeq203$MeV for two flavors of Wilson-Clover fermions with
a zero-temperature pion mass of $\simeq270$\,MeV. Exploiting exact sum rules
and applying a phenomenologically motivated ansatz allows us to determine the spectral function
 $\rho(\omega,T)$ via a fit to the lattice correlation function data. From these results we estimate the
electrical conductivity across the deconfinement phase transition via
a Kubo formula and find evidence for the dissociation of the $\rho$ meson
by resolving its spectral weight at the available temperatures. 
We  also apply the Backus-Gilbert method as a model-independent approach to this
problem. At any given frequency, it yields a local weighted average of the true
spectral function. We use this method to compare kinetic theory
predictions and previously published phenomenological spectral
functions to our lattice study.
\end{abstract}

\maketitle
%%%%%%%%%%%%%%%%%%%%%%%%%%%%%%%%%%%%%%%%%%%%%%%%%%%%%%%%%%%%%%%%
%%%%%%%%%%%%%%%%%%%%%%%%%%%%%%%%%%%%%%%%%%%%%%%%%%%%%%%%%%%%%%%%

%%%%%%%%%%%%%%%%%%%%%%%%%%%%%%%%%%%%%%%%%%%%%%%%%%%%%%%%%%%%%%%%
%%%%%%%%%%%%%%%%%%%%%%%%%%%%%%%%%%%%%%%%%%%%%%%%%%%%%%%%%%%%%%%%
\section{Introduction}

Strongly interacting matter at temperatures below 100MeV is thought to
consist of hadronic quasiparticles to a good approximation.  At
sufficiently high temperatures, perhaps of several GeV, one expects
quarks and gluons to be the relevant degrees of freedom instead. 
The stages between the two regimes are the subject of vast
experimental programs based on colliding heavy ions and of many
theoretical studies (see for instance \cite{Brambilla:2014jmp},
chapter D). The broad interest partly stems from
the fact that matter in the early universe underwent a thermal
transition between the two phases. This transition is known from
lattice simulations to be a continuous crossover rather than a
sharp phase transition~\cite{Aoki:2006we}.

% \cite{Geurts:2012rv}.

Hard probes, observables
which, although produced in the thermal medium, immediately decouple
from it, are especially interesting in this context. 
With no subsequent interaction after their production, they
are among the best observables to study the thermal medium both in
experiment~\cite{Geurts:2012rv} and theory.  For a quark gluon plasma a good probe 
is the production rate of lepton pairs, since they
experience only electroweak interactions after they are created and
act as essentially freely propagating particles. At the same time the
underlying production mechanism is highly complicated due to the
strong interactions of the constituents of the thermal system.  The
dilepton production rates for a system in thermal equilibrium are
proportional to the thermal vector spectral function. The latter
encodes the information on the excitation spectrum and transport
processes in the medium (see for instance~\cite{Meyer:2011gj}).

The hadron resonance gas (HRG) model, which is based on the idea that
the static properties of strongly interacting matter in the
low-temperature phase are given by the sum of the contributions of
non-interacting hadron species, gives an economic description of
particle yields in heavy-ion collisions (see the recent
\cite{BraunMunzinger:2011ta}, \cite{Stachel:2013zma} and references
therein) and provides a good estimate of the pressure and charge
fluctuations determined in lattice
calculations~\cite{Bazavov:2012jq,Borsanyi:2013bia,Borsanyi:2011sw}.
However, the thermal quasiparticles could have different properties,
for instance different dispersion
relations~\cite{Brandt:2014qqa,Brandt:2015sxa}, from the hadrons at
$T=0$. One of our goals here is to probe to what extent the transport
of charge can be described by the HRG model.

In a strongly coupled regime, no general computational method to
address thermal real-time phenomena is known. Interesting insights
have been gained via the gauge/gravity correspondence (see for
instance~\cite{CasalderreySolana:2011us}). Lattice QCD provides a way
to compute correlation functions in the Matsubara (or Euclidean)
formalism. These correlation functions are related to spectral
functions $\rho(\omega,T)$ via dispersion relations, as a consequence
they provide information on the latter. However, obtaining information
that is more local in the frequency $\omega$ than $\Delta\omega\sim T$
requires highly accurate calculations of the Euclidean correlation
functions. In recent years, the quality of this type of lattice
calculations has steadily
improved~\cite{Aarts:2007wj,Meyer:2007ic,Francis:2011bt,Ding:2010ga,Ding:2013qw,Ding:2014dua,Amato:2013naa,Brandt:2012jc}.

In the following we present a study of isovector vector correlation
functions across the deconfinement phase transition on large isotropic
lattice ensembles using Wilson-Clover fermions. After presenting the
theoretical background to our calculations (section \ref{sec:theory})
and our numerical framework (section \ref{sec:num}), we apply an
ansatz for the spectral functions and fit its parameters to the
lattice data (section \ref{sec:fitansatz}). Thereby we estimate the
electrical conductivity and the spectral weight of the $\rho$ meson
for a range of temperatures. In section \ref{sec:BG}, we apply the
model-independent Backus-Gilbert method \cite{Backus:1968bg} to obtain
a `filtered' spectral function, which is a local average of the
genuine spectral function around a given frequency. In this way we
obtain a means to directly compare HRG predictions and previously
published phenomenological spectral functions \cite{Hohler:2013eba} with lattice data
(section \ref{sec:compa}). The phenomenological implications of our
calculation are discussed in section \ref{sec:implic}. The article
ends with a summary of the lessons we learnt from the lattice
calculations and an outlook on how further insight could be gained on
the thermal spectral functions.

%%%%%%%%%%%%%%%%%%%%%%%%%%%%%%%%%%%%%%%%%%%%%%%%%%%%%%%%%%%%%%%%
%%%%%%%%%%%%%%%%%%%%%%%%%%%%%%%%%%%%%%%%%%%%%%%%%%%%%%%%%%%%%%%%
\section{Definitions and theory predictions\la{sec:theory}}

In this section we collect the necessary definitions and review (mostly known) facts
about correlation functions and the associated spectral functions of the vector current.

\subsection{Definitions}
Our primary observables are the Euclidean-time correlators of the conserved vector current,
\ba
G_{\mu\nu}(\tau,T) &=&   \int d^3x \; \< J_\mu(\tau,\vec x)\;J_\nu(0)^\dagger\>\,.
\ea
The expectation values are taken
with respect to the equilibrium density matrix $e^{-\beta H}/Z(\beta)$, where
$\beta\equiv 1/T$ is the inverse temperature, and the Euclidean-time evolution of operators is given 
by ${\cal O}(\tau)=e^{H\tau}{\cal O}e^{-H\tau}$.
For now we leave the flavor structure of the current unspecified and return to this aspect
in the next subsection. We define the quark number susceptibility as
\be
\chi_s = \beta \int d^3x \;\< J_0(\tau,\vec x) J_0(0)\>, \qquad \tau\neq 0.
\label{eq:chis}
\ee
The Euclidean correlators have the spectral representation
\be\la{eq:coshK}
G_{\mu\nu}(\tau,T)  \stackrel{\mu=\nu}{=} \int_0^\infty \frac{d\omega}{2\pi} \; \rho_{\mu\nu}(\omega,T) \;
\frac{\cosh[\omega(\beta/2-\tau)]}{\sinh(\omega\beta/2)}\;,
\ee
where the spectral function is defined (for a general spatial momentum $\vec p$) as
\be
\rho_{\mu\nu}(\omega,\vec p,T) = \int dt d^3x\,e^{i\omega\cdot t- i\vec p\cdot \vec x}
\Big\langle  [J_\mu(t,\vec x),J_\nu(0)^\dagger ]\Big\rangle~~,
\quad {\cal O}(t) \equiv e^{iHt}{\cal O} e^{-iHt}.
\ee

For a given function $\rho(\omega,T')$, the reconstructed correlator is defined as
\be 
G_{\rm rec}(\tau,T;T') {\equiv} \int_0^\infty \frac{d\omega}{2\pi}\; \rho(\omega,T')\;
\frac{\cosh[\omega(\beta/2-\tau)]}{\sinh( \omega \beta/2)} \,.
\la{eq:Grec1-main}
\ee
It can be interpreted as the Euclidean correlator that would be
realized at temperature $T$, if the spectral function did not change between
temperature $T$ and $T'$.  For $T'=0$ it can be directly obtained from the
zero-temperature Euclidean correlator via~\cite{Meyer:2010ii}
\be
G_{\rm rec}(\tau,T) \equiv G_{\rm rec}(\tau,T;0) = 
\sum_{m\in\mathbb{ Z}} G(|\tau+m\beta|,T=0).
\la{eq:Grec2-main}
\ee

\subsection{On the flavor structure of the current}

Unless otherwise stated, we  consider two-point functions
of the isospin current\footnote{Note that in this normalization, all
  two-point functions of the current are a factor 2 larger than the
  isovector contributions to the two-point functions of the
  electromagnetic current.}
\be
J_\mu(x)\equiv
\frac{1}{\sqrt{2}}\left(\bar u(x)\gamma_\mu u(x) - \bar d(x)\gamma_\mu d(x)\right)
\ee
in QCD with two light degenerate flavors of quarks, and the associated spectral function
$\rho_{\mu\nu}(\omega,T)$.

Many phenomenological observables are related to the electromagnetic current
\be
J_\mu^{\rm em} = \sum_f Q_f \bar q_f \gamma_\mu q_f 
\ee
with $Q_u=2/3$, $Q_d=-1/3$ etc.\ and the associated spectral function
$\rho_{\mu\nu}^{\rm em}(\omega,\vec p,T)$.  In two-flavor QCD,
we can decompose the electromagnetic current into the following sum of 
isovector and isoscalar components,
\be
J_\mu^{\rm em} = \frac{1}{\sqrt{2}} J_\mu + \frac{1}{2}J_\mu^{\rm B},
\ee
where
\be
\qquad J_\mu^{\rm B} = \frac{1}{3} \left(\bar u(x)\gamma_\mu u(x) + \bar d(x)\gamma_\mu d(x)\right)
\ee
is the baryon current. Since in this work we compute only the
two-point function of $J_\mu$, the question poses itself as to whether
the two-point function of $J_\mu^{\rm em}$ can be approximated with
the available information. In the vacuum, this question has been
addressed recently~\cite{Francis:2013fzp,Francis:2014hoa}. It turns out that, as far
as the spectral function is concerned, the appropriate approximation
depends on the frequency. At sufficiently high frequencies, a decomposition 
of the two-point function of $J_\mu^{\rm em}$ in terms of a quark-line connected contribution
and a disconnected contribution is most useful, since the latter is of order
$\alpha_s^3$ in perturbative QCD. When the quark-line disconnected contribution
to the two-point function of $J_\mu^{\rm em}$ can be neglected, we have 
\be\la{eq:appro5/9}
\rho_{\mu\nu}^{\rm em}(\omega,\vec 0,T)|_{n_f=2} \simeq C_{\rm em} \rho_{\mu\nu}(\omega,T),
\qquad C_{\rm{em}} =\sum_{f=u,d} Q_f^2=5/9.
\ee
On the other hand, isospin symmetry implies that the two-point
function of $J_\mu^{\rm em}$ is given by the sum of the two-point
function of $\frac{1}{\sqrt{2}}J_\mu$ and of the two-point function of $\frac{1}{2}J_\mu^{\rm B}$.
For $\omega\lesssim m_\omega$, where $m_\omega$ is the mass of the
(isoscalar vector) $\omega$ meson, the isoscalar two-point function
is negligible compared to the isovector contribution; in fact, the
corresponding contribution to the spectral function is exactly zero
for $0\leq \omega\leq 3m_\pi$. Therefore, in this low-frequency regime the appropriate
approximation is 
\be\la{eq:appro1/2}
\rho_{\mu\nu}^{\rm em}(\omega,\vec 0,T) ~\simeq~ \frac{1}{2} \rho_{\mu\nu}(\omega,T),
\qquad 0\leq \omega\lesssim m_\omega.
\ee

One may ask whether any of the two presented approximations still
holds at non-zero temperature.  At sufficiently high temperatures,
when the relevant degrees of freedom are quarks and gluons, we expect
that the approximation given in Eq.\ (\ref{eq:appro5/9}) is applicable at all
frequencies.  In the low-temperature phase, we expect the same
approximation to hold as for zero temperature, namely
\eq(\ref{eq:appro1/2}) for $\omega\lesssim m_\omega$ and
\eq(\ref{eq:appro5/9}) at higher frequencies.  This expectation is
based in particular on the observation that only baryons contribute to
the transport peak in the spectral function associated with the
current $J_\mu^{\rm B}$ (see section \ref{sec:kin} below); this transport peak therefore carries far
less spectral weight than the transport peak in the isovector channel,
since the latter receives contributions from pions, $\rho$ mesons etc.

\subsection{General properties of spectral functions and their relation to observables}

In the thermodynamic limit, the subtracted vector spectral function obeys the exact sum rule
(see~\cite{Bernecker:2011gh} sec.\ 3.2),
\be\la{eq:sr}
\int_{-\infty}^\infty \frac{d\omega}{\omega} \; \Delta\rho(\omega,T) = 0,
\qquad \qquad \Delta\rho(\omega,T) \equiv  \rho_{ii}(\omega,T)-\rho_{ii}(\omega,0).
\ee

The diffusion constant $D$ 
is given by a Kubo formula in terms of the low-frequency behavior of the spectral function,
\be
D = \frac{1}{6\chi_s}\;\lim_{\omega\to0} \frac{\rho_{ii}(\omega,T)}{\omega}.
\la{eq:kubo}
\ee
The electrical conductivity in two flavor QCD is given by 
\be
\la{eq:sigD}
\sigma_{\rm el} = \frac{1}{6}\;\lim_{\omega\to0} \frac{\rho^{\rm em}_{ii}(\omega,\vec 0,T)}{\omega}
  \simeq C_{em} D\chi_s~. 
\ee
In the above equation, disconnected diagrams have been neglected.
More generally, the electromagnetic spectral function determines the differential production rate of lepton-antilepton pairs
per unit volume of the thermal system,
\be\la{eq:dilrate}
\frac{dN_{l^+l^-}}{d\omega d^3p}= C_{\rm em}~ \frac{\alpha_{\rm em}^2}{6\pi^3}~
\frac{\rho^{\rm em}_{\mu\mu}(\omega,\vec p,T)}{(\omega^2-\vec p^2)(e^{\omega/T}-1)}~~~.
\ee

\subsection{Non-interacting limits of the spectral functions}

\noindent There are two simple limits, in which the spectral function can be determined
analytically.  For non-interacting massive quarks in the fundamental
representation of the SU($N_c$) color group, the vector spectral
function is given by\footnote{At large frequencies the radiative corrections
$(1+\alpha_s/\pi+\dots)$ to the coefficient of the $\omega^2$ term are
temperature independent and known to order
$\alpha_s^4$~\cite{Baikov:2012zm} (for quark mass effects in the
vacuum, see~\cite{Chetyrkin:1997qi}).}
\be\la{eq:SFfree}
\rho_{ii}(\omega,T) =  2\pi \chi_s \<v^2\>\omega \delta(\omega) 
%  \\ && 
+ \frac{N_c}{2\pi} \theta(\omega-2m) \left[1- \frac{4m^2}{\omega^2}  \right]^{\frac{1}{2}}
\left[1+   \frac{2m^2}{\omega^2}  \right] \omega^2 \tanh({\omega\beta}/{4}).
% \nonumber
\ee
The sum rule Eq.~(\ref{eq:sr}) is verified by this spectral function.
On the other hand, at low  temperatures and at low  frequency, we expect
pions to be the relevant degrees of freedom. In the theory of non-interacting pions,
the third isospin component of the vector current is represented by
$J_\mu= \sqrt{2}(\pi^1 \partial_\mu \pi^2 - \pi^2 \partial_\mu \pi^1)
= \sqrt{2}\,i(\pi^+\partial_\mu\pi^- - \pi^- \partial_\mu\pi^+)$.
The spectral function reads
\be
\rho_{ii}(\omega,T) = 2\pi \chi_s \<v^2\>\omega \delta(\omega)
+ \frac{1}{4\pi} \theta(\omega-2m_\pi) \left[1-\frac{4m_\pi^2}{\omega^2}\right]^{\frac{3}{2}} 
\frac{\omega^2}{\tanh(\omega\beta/4)}.
\ee
For free quarks, the susceptibility $\chi_s$ and the mean squared transport
velocity $\<v^2\>$ are given by,
\ba\la{eq:chi_quark}
\chi_s &=& {4N_c\beta}\int \frac{d^3\vec p}{(2\pi)^3} \; f^{_F}_{\vec p} (1-f^{_F}_{\vec p})~~,
\\\la{eq:chi_quark2}
\chi_s \<v^2\> &=& {4N_c\beta}\int \frac{d^3\vec p}{(2\pi)^3}\; f^{_F}_{\vec p} (1-f^{_F}_{\vec p})\;
               \frac{\vec p^2}{E_{\vec p}^2}~~,
\ea
while for free pions the expressions are\footnote{Note, the pion contribution to the 
susceptibility of the electric charge $Q$
is $1/2$ times the right-hand side of \eq(\ref{eq:chi_pion}).}
\ba\la{eq:chi_pion}
\chi_s &=& 4\beta \int \frac{d^3\vec p}{(2\pi)^3} \; f^{_B}_{\vec p} (1+f^{_B}_{\vec p})~~,
\\\la{eq:chi_pion2}
\chi_s \<v^2\> &=& 4\beta \int \frac{d^3\vec p}{(2\pi)^3} \; f^{_B}_{\vec p} (1+f^{_B}_{\vec p})\, 
\frac{\vec p^2}{E_{\vec p}^2}~~.
\ea
The energies in Eqs.~(\ref{eq:chi_quark})-(\ref{eq:chi_pion2}) are respectively the free energies of quarks
$E_{\vec p} =\sqrt{\vec p^2+m^2}$ and pions $E_{\vec p} = \sqrt{\vec p^2+m_\pi^2}$,
while $f^{_{B/F}}_{\vec p}= 1/[e^{\beta E_{\vec p}}\mp 1]$ are the Bose-Einstein and
Fermi-Dirac distributions. The mean square velocity is unity for massless particles
and $3T/m$ for non-relativistic particles.
The susceptibility of free particles has simple expressions in the high-temperature limit
\ba
\chi_s \stackrel{T\to\infty}{=} \left\{\begin{array}{c}
({N_c}/3)\,T^2 \\
({2}/{3}) \,T^2
\end{array}\right.,
\ea
and in the low-temperature limit
\ba
\chi_s \stackrel{T\to0}{=} \left\{\begin{array}{c}
 \frac{\sqrt{2}}{\pi^{3/2}}\,  T^{1/2} m^{3/2} \,e^{-\beta m}
 \\
N_c \frac{\sqrt{2}}{\pi^{3/2}}\,  T^{1/2} m^{3/2} \,e^{-\beta m}
\end{array}\right.,
\ea
where the upper cases correspond to quarks and the lower ones to pions.

Beyond the strict non-interacting theory, weak coupling kinetic theory
predicts the presence of a narrow transport peak in the spectral
function at $\omega=0$, whose width and height are related to the
properties of the quasi-particles. Introducing a separation scale
$\Lambda$ between the transport time scale and the thermal time-scale,
the area under the transport peak is, to leading order, preserved by
the interactions~\cite{Petreczky:2005nh},
\be\la{eq:IntLbda}
\mathcal{A}(\Lambda)=\int_{-\Lambda}^\Lambda \frac{d\omega}{2\pi} \frac{\rho_{ii}(\omega,T)}{\omega} = 
\chi_s \<v^2\>~~,
\ee
and the width of the transport peak becomes finite.

\subsection{Kinetic theory and the hadron resonance gas\la{sec:kin}}

We now explore the assumption that the thermal system in the low-temperature phase
is well approximated by a gas of weakly interacting hadrons.
In~\cite{Brandt:2012jc}, we estimated that the area under the $\rho$ resonance peak in the 
vacuum spectral function $\rho_{ii}(\omega,0)$ was 
\be\la{eq:114}
\int_0^{1{\rm GeV}} \frac{d\omega}{\pi} \frac{\rho_{ii}(\omega,0)}{\omega} = 0.114\,{\rm GeV}^2.
\ee
As we shall see shortly, the area under the transport peak represents
no more than a $\sim 5\%$ effect in comparison to the spectral weight
of the $\rho$ meson\footnote{From the point of view of large-$N_c$ counting
rules, this is not surprising, since the spectral weight of the $\rho$
meson is of order $N_c$, while the area under the transport peak is of
order $N_c^0$.}.

The hadron resonance gas describes the thermodynamic properties and
the quark number susceptibilities of the low-temperature phase rather
well. Assuming a transport peak exists in $\rho_{ii}(\omega,T)$, it is
interesting to ask whether the area under the peak is consistent with this picture.
It is straightforward to compute the sum of the mesonic and baryonic contributions,
e.g.\ $\chi_s\<v^2\> = (\chi_s \<v^2\>)_{\rm mesons} + (\chi_s \<v^2\>)_{\rm baryons}$,
\ba\la{eq:hrgmes}
(\chi_s \<v^2\>)_{\rm mesons} &=& \frac{2\beta}{3} \sum_{\rm multiplets} (2J+1) I(I+1)(2I+1) 
\int\frac{d^3\vec p}{(2\pi)^3} f^{_B}_{\vec p} (1+f^{_B}_{\vec p})\, \frac{\vec p^2}{E_{\vec p}^2}~,
\\\la{eq:hrgbar}
 ( \chi_s \<v^2\> )_{\rm baryons} 
&=& \frac{4\beta}{3} \sum_{\rm multiplets} (2J+1) I(I+1)(2I+1) 
\int\frac{d^3\vec p}{(2\pi)^3} f^{_F}_{\vec p} (1-f^{_F}_{\vec p}) \, \frac{\vec p^2}{E_{\vec p}^2}~.
\ea
The sums are carried out over the multiplets\footnote{The additional
  factor of 2 present in the baryon case accounts for antiparticles;
  in the meson case, the antiparticles are already included by summing
  $\sum_{I_3=-I}^I I_3^2 = \frac{1}{3} I(I+1) (2I+1)$.}  of spin $J$
and isospin $I$.  The corresponding expressions for $(\chi_s)_{\rm
  mesons}$ and $(\chi_s)_{\rm baryons}$ are identical respectively to
Eqs.~(\ref{eq:hrgmes}) and (\ref{eq:hrgbar}), except for the absence
of the factor $\frac{\vec p^2}{E_{\vec p}^2}$ in the
integrand. Expanding the factor $f(1\pm f)$ in a Taylor series in $e^{-\beta
  E_{\vec p}}$, one can express it in terms of a sum of Bessel
functions (see for instance the expressions in~\cite{Bazavov:2012jq}).
At physical quark masses in two-flavor QCD, we thus obtain the HRG estimates 
(more details on our implementation of the HRG model are given in section \ref{sec:vsq})
\be
\chi_s \<v^2\> / {\rm GeV}^2 = \left\{ \begin{array}{c@{\qquad}c}
\,0.00236 &  T=100\,{\rm MeV} \\
0.0042 &  T=120\,{\rm MeV} \\
0.0069 & T=140\,{\rm MeV} \\
0.0111 & T=160\,{\rm MeV} \\
 \end{array} \right.
\la{eq:chisv^2HRG}
\ee
The pions dominate this quantity up to quite high temperatures: for instance, at $T=140\,{\rm MeV}$, 
they still contribute $90\%$ of the total value.
We remark that combining the exact sum rule Eq.~(\ref{eq:sr}) with the kinetic sum rule Eq.~(\ref{eq:IntLbda}), we have the property
\be
\int_{\Lambda}^\infty \frac{d\omega}{\pi} \frac{\Delta\rho(\omega,T)}{\omega} = -\chi_s \<v^2\>.
\ee
Since the values in \eq(\ref{eq:chisv^2HRG}) are much smaller than the
area under the zero-temperature spectral function up to
$\omega=1{\rm GeV}$ given in Eq.~(\ref{eq:114}), the spectral weight above
the threshold $\omega=2m_\pi$ is not  substantially modified by thermal effects as long as the
HRG remains a good approximation for evaluating $\chi_s \<v^2\>$. The HRG model is in that sense self-consistent.
% Once $\chi_s$ reaches about $0.06{\rm GeV}^2$, the spectral weight above the threshold $\omega=2m_\pi$

%%%%%%%%%%%%%%%%%%%%%%%%%%%%%%%%%%%%%%%%%%%%%%%%%%%%%%%%%%%%%%%%
%%%%%%%%%%%%%%%%%%%%%%%%%%%%%%%%%%%%%%%%%%%%%%%%%%%%%%%%%%%%%%%%
\section{Numerical Setup}
\la{sec:num}

We present a study of thermal isovector vector two-point functions for
temperatures $T/T_c=0.16$, 0.8, 1.0, 1.25 and 1.67 in two-flavor QCD.  We
vary the temperature by increasing the number of lattice sites in the
time directions at fixed bare parameters, a method known as the
`fixed-scale' approach. The mass of the degenerate quark doublet
corresponds to a zero-temperature pion mass of
$\simeq270$MeV~\cite{Capitani:2015sba}. The data presented here is thus a
natural extension of the calculation at $T/T_c=1.25$ presented
in~\cite{Brandt:2012jc}.  The gauge action is the standard Wilson
plaquette action \cite{Wilson:1974sk}, while the fermions are
implemented via the O($a$) improved Wilson discretization with the
non-perturbatively determined clover coefficient $c_{\rm sw}$
\cite{Jansen:1998mx}.  All configurations are generated using the
MP-HMC algorithm~\cite{Hasenbusch:2001ne,Hasenbusch:2002ai} in the
implementation of Marinkovic and Schaefer~\cite{Marinkovic:2010eg}
based on L\"uscher's DD-HMC package~\cite{Luscher:2005rx}.

%%%%%%%%%%%%%%%%%%%%%%%%%%%%%%%%%%%%%%%%%%%%
\begin{table}[h!]
\centering
 % Give a unique label
% For LaTeX tables use
\begin{tabular}{llc|cll}
\hline\hline\noalign{\smallskip}
$6/g_0^2$ & 5.50 		&&& $m_\pi$[MeV]   &  $\simeq270$\\
$\kappa$ & 0.13671 	&&& $Z_V$ & 0.768(5)\\
$c_{SW}$ & 1.751496 	&&& $a[\textrm{fm}]$ & 0.0486(4)(5) \\
\noalign{\smallskip}\hline\hline\noalign{\smallskip}
$N_s^3\times N_\tau$ & $64^3 \times 128$ &&& $N^{\rm vac}_{\rm conf}$ & 137\\
$T^{\rm vac}$[MeV] & $32(6)$&&&$N^{\rm vac}_{\rm src}$ & 16 \\
\noalign{\smallskip}\hline\hline\noalign{\smallskip}
$N_s^3\times N_\tau$ & $64^3 \times 24$ &&&$N_s^3\times N_\tau$ & $64^3 \times 20$\\
$T$[MeV] & $169(3)$ 					 &&&$T$[MeV] & $203(4)$\\
$N_{\rm conf}$ & 360 						 &&&$N_{\rm conf}$ & 311 \\ 
$N_{\rm src}$ & 64 							 &&&$N_{\rm src}$ & 64 \\
\noalign{\smallskip}\hline\noalign{\smallskip}
$N_s^3\times N_\tau$ & $64^3 \times 16$ &&& $N_s^3\times N_\tau$ & $64^3 \times 12$ \\
$T$[MeV] & $254(4)$ 					 &&& $T$[MeV] & $338(5)$\\
$N_{\rm conf}$ & 313 						 &&& $N_{\rm conf}$ & 262\\ 
$N_{\rm src}$ & 65 							 &&&$N_{\rm src}$ & 65 \\
\noalign{\smallskip}\hline\hline
% \noalign{\smallskip}\hline
\end{tabular}
\caption{{The top left block shows the bare lattice parameters, see
    also~\cite{Fritzsch:2012wq,Brandt:2012jc,Brandt:2012sk}.  The top
    right block summarizes the pion mass, the vector renormalization
    constant~\cite{DellaMorte:2005rd} and the lattice
    spacing~\cite{Fritzsch:2012wq}.  The middle and bottom blocks
    contain more specific information on the $N_\tau=128, 24, 20, 16$
    and $12$ lattice calculations, such as their corresponding
    temperatures, their number of configurations and the number of
    source positions used to calculate the correlation functions.}}
\label{tab:latticepar}      
\end{table}
%%%%%%%%%%%%%%%%%%%%%%%%%%%%%%%%%%%%%%%%%%%%

The spatial
size of the ensembles is $N_s=64$ with periodic boundary conditions 
and the temporal extents are
$N_\tau=128, 24, 20, 16$ and 12. The $N_\tau=16$ ensemble was first
presented in~\cite{Brandt:2012jc} and has subsequently been analyzed
in \cite{Brandt:2013faa,Brandt:2013fg,Brandt:2014uda}.  All ensembles
are calculated at fixed bare parameters, whereby the lattice spacing
is $a=0.0486(4)(5)$fm~\cite{Fritzsch:2012wq} and $m_\pi L = 4.2$. With
$T=1/(N_\tau a)$ the ensembles correspond to the temperatures
$T=169(3)$MeV $(N_\tau=24)$, $T=203(4)$MeV $(N_\tau=20)$,
$T=254(4)$MeV $(N_\tau=16)$ and $T=338(5)$MeV $(N_\tau=12)$. Based on
preliminary results on the pseudo-critical temperature $T_c$ of the
crossover from the hadronic to the high-temperature
phase~\cite{Brandt:2012sk,Brandt:2013mba}, the temperatures can also
be expressed in terms of $T_c$ by $T/T_c\approx 0.8, 1.0, 1.25$ and
1.67.  In addition, we have updated our vacuum correlation function at
$N_\tau=128$, i.e. $T=32(6)$MeV, by significantly increasing the statistics.
We refer to this as the `vacuum' ensemble.

In contrast to \cite{Brandt:2012jc}, we implement the vector
correlation function on the lattice 
as a mixed correlator between the local and the conserved current,
\ba \la{eq:gcorr}
G_{ii}^{\rm bare}(\tau,g_0,T) &=& - a^3 \sum_{i=1}^{3} \sum_{\vec x} \,\< J^c_i(\tau,\vec x) J^\ell_i({0}) \>,
\\
G_{00}^{\rm bare}(\tau,g_0,T) &=& - a^3 \sum_{\vec x} \,\< J^c_0(\tau,\vec x) J^\ell_0({0}) \>,
\ea
where
\ba
J_\mu^\ell({x}) &=& \frac{1}{\sqrt{2}}\bar q({x}) \gamma_\mu {\tau^3} q({x}),
\\
\la{eq:Jdef}
J_\mu^c ( {x} ) &=& \frac{1}{2\sqrt{2}} \Big(\bar q ( {x} + a\hat\mu ) ( 1 + \gamma_\mu ) U_\mu^\dagger ( {x} ) {\tau^3} q ( x ) 
 - \bar q ( {x} )( 1 - \gamma_\mu ) U_\mu ( {x} ) {\tau^3} q ( {x} + a\hat\mu ) \Big).
\ea
Here $U_\mu(x)$ are the gauge links, $q$ represents a doublet of mass-degenerate quark fields and
$\tau^3$ the diagonal Pauli matrix acting on the flavor indices. The
doublet can be understood as the (u,\,d) quarks and are treated fully
dynamically in this calculation. To achieve a precision at the
$\lesssim 1\%$ level we supplement the source at position
$x_{\rm src}=(0,0,0,0)$ with additional $N_{\rm src}=64$ randomly
chosen source positions in the lattice four-volume in order to increase statistics
by exploiting the translational invariance of the system. 
The correlators are renormalized via
\be
G_{\mu\nu}(\tau,T)= Z_V(g_0)\;  G_{\mu\nu}^{\rm bare}(\tau,g_0,T) 
\ee
with the non-perturbative value of
$Z_V=0.768(5)$~\cite{DellaMorte:2005rd}.  Here, our primary goal is to
carry out the analysis on a single lattice spacing.  We therefore have
not included O($a$) contributions from the improvement term
proportional to the derivative of the antisymmetric tensor
operator~\cite{Luscher:1996sc,Sint:1997jx}. Also a quark-mass
dependent improvement term $(1+b_V(g_0)am_q)$~\cite{Sint:1997jx} was
neglected. These contributions should eventually be included to ensure
a smooth scaling behavior as the continuum limit is taken.  The
parameters of the lattice ensembles are collected in
Tab.~\ref{tab:latticepar}. In addition all results for the
correlators are given in Tab.~\ref{tab:vacdat} and
Tab.~\ref{tab:thermdat} in Appendix~\ref{app:data}. The covariance
matrices of these data sets are provided online on the arXiv \cite{arxiv}.

The reconstructed correlator $G_{\rm rec}(\tau,T)$ 
can be straightforwardly computed when the ratio of vacuum and target temperature 
$N_\tau^{\rm vac}/N_\tau^{\rm target}$ is an integer, which is the case for the 
$N_\tau^{\rm target}=16$ ensemble, as the ratio is $N_\tau^{\rm
  vac}/N_\tau^{\rm target}=8$.
For the other temperatures $N_\tau^{\rm target}=24,20,12$, this strategy is not immediately
applicable.  One option is to fold the vacuum
data around $m\cdot N_\tau^{\rm target}$, following the recipe of
Eq.~(\ref{eq:Grec2-main}), until the maximum of possible foliations is
reached.  In this case we assume that the large distance foliations of
the vacuum correlator in Eq.~(\ref{eq:Grec2-main}) provide a negligible
contribution to the overall result.  An alternative way is to first use a
cosh-fit beyond some distance $\tau^{\rm cut}$ to extend the vacuum
correlator data to all $\tau\in\mathbb{R}$. 
In the following we set $\tau^{\rm cut}/a=36$.  Checking these
two methods on the exactly computable $N_\tau^{\rm target}=16$
ensemble, we observe that both of them agree within the statistical errors of the corresponding thermal data.

%%%%%%%%%%%%%%%%%%%%%%%%%%%%%%%%%%%%%%%%%%%%%%%%%%%%%%%%%%%%%%%%
%%%%%%%%%%%%%%%%%%%%%%%%%%%%%%%%%%%%%%%%%%%%%%%%%%%%%%%%%%%%%%%%
\section{Vector spectral functions in two-flavor QCD}

With precision data available, we note that two approaches have been
widely adopted to reconstruct spectral functions from lattice
correlators. The first is the maximum entropy method (MEM)
\cite{Asakawa:2000tr,Wetzorke:2001dk,Aarts:2007wj,Burnier:2013nla}. Here
Bayes' theorem is invoked to determine the most probable spectral
function given the data on the one hand and a so-called default model
on the other. Standard algorithms maximize an entropy term to
determine the coefficients of a set of transformed basis functions
$\mathcal{L}(B(k))$ to obtain an approximation of the spectral
function based on the correlator data.  Since the inverse
transformation of the basis functions is known, one thereby obtains
also an estimate of the spectral function. The main caveat in this
method is the dependence of the results on the default model and, to
some extent, on the basis of functions in which the spectral function
is expanded. Alternatively, instead of fixing the basis functions and
determining their weights, one can define an ansatz $F(a_k,\omega)$,
with parameters $a_k$, for the spectral function
\cite{Karsch:1986cq,Ding:2010ga,Brandt:2012jc}.  Naturally, the ansatz
introduces a model-dependence and has to be justified.  If the fit is
successful, one obtains, as in the case of the MEM, 
a spectral function which describes the lattice data within its statistical uncertainty.

In addition to updating our recent analysis using the fit ansatz
approach, we also apply the Backus-Gilbert
method \cite{Brandt:2015sxa}. The idea of the method is
to determine a local average of the spectral function around a given
value of $\omega$ without parametrizing it in any particular way\footnote{We note that an alternative model independent
  reconstruction procedure based on Cuniberti's method was discussed
  in \cite{Burnier:2011jq,Burnier:2012ts}.}. The
weighting factor of the average is called the resolution function.

%%%%%%%%%%%%%%%%%%%%%%%%%%%%%%%%%%%%%%%%%%%%%%%%%%%%%%%%%%%%%%%%
%%%%%%%%%%%%%%%%%%%%%%%%%%%%%%%%%%%%%%%%%%%%%%%%%%%%%%%%%%%%%%%%
\subsection{Fit ansatz for the spectral function}
\la{sec:fitansatz}

In the first step of this analysis, we reconstruct the vector meson
thermal spectral functions using the ansatz approach. Specifically we
determine the vacuum and thermal spectral functions from simultaneously 
fitting all available correlator data $G(\tau,T)$ and enforcing the exact 
sum rule of \eq(\ref{eq:sr}). 

\subsubsection{Combined fit for the vacuum and thermal spectral functions}

To motivate a sensible ansatz for the spectral function on
our `$T=0$' ensemble, we first note that the transport contribution is
absent in the vacuum. Given the ensemble parameters, we describe the $\rho$ meson contribution 
by a $\delta$-function.  The large
frequency behavior is parametrized by a sharp threshold to the
perturbative behavior,
\be\la{eq:vacspf}
\frac{\rho_{V}(\omega,T\simeq 0)}{2\pi}
= a_{V}\delta(\omega-m_V) + \frac{3\kappa_0}{4\pi^2}\Theta(\omega-\Omega_0)\,\omega^2\,\tanh\Big(\frac{\omega\beta_0}{4}\Big),~~~(\omega\geq 0).
\ee
Based on similar considerations, we define a set of thermal spectral function models for $(\omega\geq 0)$.

As initial setup, we propose that the thermal spectral function consists of a bound state $\delta$-peak, a continuum threshold, a simple $\delta$-peak for the transport and an OPE inspired term ($\sim 1/\omega^{2}$), which is denoted (Mod. 1) in the following. The ansatz reads:
\be\la{eq:thespf1}
\frac{\rho_{\rm Mod. 1}(\omega,T)}{2\pi}
= A_{\rm T}\omega\delta(\omega) + a_{\rm T}\delta(\omega-m_{\rm T}) + \frac{3\kappa_0}{4\pi^2}\Theta(\omega-\Omega_{\rm T})\,\omega^2\tanh\Big(\frac{\omega\beta_{\rm T}}{4}\Big) + \frac{3\kappa_{\rm O}}{4\pi^2}\Theta(\omega-\Omega_{\rm O})\,\frac{1}{\omega^2} .
\ee
To permit statements on the electrical conductivity, we further introduce a Breit-Wigner type transport peak, while keeping the $\delta$-function for the $\rho$ meson.
In addition to introducing a finite width for the transport region, we allow that $\kappa_0$ splits into two components below the vacuum threshold at $\Omega_0$:
\be\la{eq:thespf2b}
\kappa_0 \rightarrow \tilde\kappa_0(\kappa_0,\kappa_{1},\Omega_0,\eta)=\Big[ \kappa_0 + \kappa_{1}\Big(1-\tanh\Big(\frac{\omega}{\Omega_0\cdot\eta} \Big)^2\Big)\Big].
\ee
Our second and main model is given by (Mod. 2):
\be\la{eq:thespf2}
\frac{\rho_{\rm Mod. 2}(\omega,T)}{2\pi}
= \frac{\omega\,A_{\rm T}\,\Gamma_{\rm T}}{\pi(\Gamma_{\rm T}^2 + \omega^2)} + a_{\rm T}\delta(\omega-m_{\rm T}) + \frac{3\tilde\kappa_0}{4\pi^2}\Theta(\omega-\Omega_{\rm T})\omega^2\,\tanh\Big(\frac{\omega\beta_{\rm T}}{4}\Big)+ \frac{3\kappa_{\rm O}}{4\pi^2}\Theta(\omega-\Omega_{\rm O})\,\frac{1}{\omega^2}.
\ee
As next ingredient we incorporate the sum rule of Eq.~(\ref{eq:sr}) and find e.g.
\be\la{eq:srcorr}
A_{\rm T} = 2\Big( \frac{a_0}{m_0} - \frac{a_{\rm T}}{m_{\rm T}} - \frac{3\kappa_{\rm O}}{8\pi^2 \Omega_{\rm O}^2}
- \frac{3\kappa_0}{4\pi^2}\int_0^\infty d\omega\,\omega\,
\Big[ \Theta(\omega-\Omega_{\rm T})\tanh\Big(\frac{\omega\beta_{\rm T}}{4}\Big) - \Theta(\omega-\Omega_0)\tanh\Big(\frac{\omega\beta_0}{4}\Big)   \Big]\Big)~~,
\ee
for the first model (Mod. 1). An analogous expression for the second model is readily derived. 
The parameter $A_{\rm T}$ is eliminated from the thermal fit functions by enforcing the sum rule in a combined fit with the vacuum correlator. In the following we use the vacuum and thermal ans\"atze to fit the data of all ensembles simultaneously. In addition to eliminating $A_{\rm T}$ via Eq.~(\ref{eq:srcorr}), we enforce the following conditions:
The parameters $\kappa_0$ as well as $m_{\rm V}=m_{\rm T}$ are shared parameters for all ensembles. Without these constraints the fits lead to considerably less well determined parameters. The threshold parameters $\Omega_0$ and 
$\Omega_T$ are shared for the $N_\tau=128, 24$ ensembles, while $\Omega_T$ is set to zero for $N_\tau=16$ and 12. The free parameter $\Omega_T$ is exclusive to the $N_\tau=20$ ensemble and is set according Tab.~\ref{tab:variants}.
The OPE inspired parameter $\Omega_{\rm O}$ is set to the vacuum value $\Omega_0$. 

\begin{table}[t!]
\centering
 % Give a unique label
% For LaTeX tables use
\begin{tabular}{lccccccccccc}
\hline\hline\noalign{\smallskip}
 %& & & \multicolumn{2}{c}{bound state} & \multicolumn{2}{c}{cont. thresh.} &
%\multicolumn{2}{c}{~~~OPE} \\
%\hline\noalign{\smallskip}
Ansatz & $N_\tau$ & $T/T_c$ &~~~~& $a_T$ & $m_T$ &~~~~& $\kappa_{1}$ & $\Omega_T$ &~~~~&
$\kappa_{\rm O}$ & $\Omega_{\rm O}$~ \\
\hline\noalign{\smallskip}
(Mod. 1) & 24 & 0.80 && free & $m_V$ && - & $\Omega_0$ && free & $\Omega_0$
\\
         & 20 & 1.00 && free & $m_V$ && - & 0 && free & $\Omega_0$ \\
         & 16 & 1.25 && free & $m_V$ && - & 0 && free & $\Omega_0$ \\
         & 12 & 1.67 && free & $m_V$ && - & 0 && free & $\Omega_0$ \\
\hline\noalign{\smallskip}
(Mod. 2a) & 24 & 0.80 && free & $m_V$ && 0 & $\Omega_0$ && free & $\Omega_0$ \\
          & 20 & 1.00 && free & $m_V$ && 0 & $\Omega_0$ && free & $\Omega_0$ \\
          & 16 & 1.25 && 0 & - && free & 0 && 0 & - \\
          & 12 & 1.67 && 0 & - && free & 0 && 0 & - \\
\hline\noalign{\smallskip}
(Mod. 2b) & 24 & 0.80 && free & $m_V$ && 0 & $\Omega_0$ && free & $\Omega_0$ \\
          & 20 & 1.00 && free & $m_V$ && free & 0 && 0 & - \\
          & 16 & 1.25 && free & $m_V$ && free & 0 && 0 & - \\
          & 12 & 1.67 && free & $m_V$ && free & 0 && 0 & - \\
\hline\noalign{\smallskip}
(Mod. 2c) & 24 & 0.80 && free & $m_V$ && 0 & $\Omega_0$ && free & $\Omega_0$ \\
          & 20 & 1.00 && free & $m_V$ && free & 0 && 0 & - \\
          & 16 & 1.25 && 0 & - && free & 0 && 0 & - \\
          & 12 & 1.67 && 0 & - && free & 0 && 0 & - \\
\hline\noalign{\smallskip}
(Mod. 2d) & 24 & 0.80 && free & $m_V$ && 0 & $\Omega_0$ && free & $\Omega_0$ \\
          & 20 & 1.00 && 0 & - && free & free && 0 & - \\
          & 16 & 1.25 && 0 & - && free & 0 && 0 & - \\
          & 12 & 1.67 && 0 & - && free & 0 && 0 & - \\
\hline\hline
\end{tabular}
\caption{{Parameter setup for the ans\"atze discussed in the text. The
unlisted parameter $\Gamma_T$ is always free, while $\kappa_0$ is a parameter
which is shared for all temperatures. The parameter $A_T$ is eliminated by
the sum rule. Whenever the table contains a "-" the associated parameter
does not appear. When the entry equals one of the
zero-temperature parameters, it is a shared parameter. Note,
that the parameter $a_T$ is always constrained to $a_T>0$.}}
\label{tab:variants}      
\end{table}

We consider four variants of the second model, as listed in Tab.~\ref{tab:variants}. In the vicinity of the critical temperature, it is difficult to provide an argument for a vacuum-like or a thermal-like structure of the spectral function. By testing variants and determining the stable features we consequently obtain a more systematically reliable result. The first variant, (Mod. 2a), enforces a vacuum structure with a continuous threshold located at $\Omega_0$.
The final variant, (Mod. 2d), on the other hand leaves the threshold free and assumes a largely thermal structure. Both variants exclude thermal bound state peaks in the high temperature phase. 
The second and third variants, (Mod. 2b) and (Mod. 2c), are closely related.  Both set the threshold at zero, while allowing a thermal modification via $\kappa_{1}$, but also include a bound state peak for the $N_\tau=20$ ensemble. The difference is that (Mod. 2b) allows for bound state peaks in the high temperature phase by setting $a_T>0.0$.
This model has the most freedom to interpolate between a thermal-like and vacuum-like structure. At the same time, it results in the  largest uncertainties on the parameters.
Between the tight constraints of (Mod. 2a) or (Mod. 2d) and the relative freedom of (Mod. 2b), the remaining model (Mod. 2c) represents a trade off. Consequently, if not stated otherwise, all results shown in the following are obtained from model (Mod. 2c). \\
Throughout, the fit range chosen is $\tau/a\in[4:48]$ for the vacuum and $\tau/a\in[4:N_\tau/2]$ for the thermal ensembles. 
We tested additional models, for example also including a Breit-Wigner type peak for the $\rho$ meson. This class of models however lead to badly constrained parameters and a poor description of the data.

%%%%%%%%%%%%%%%%%%%%%%%%%%%%%%%%%%%%%%%%%%%%%%%%%%%%%%%%%%%%%%%%
\begin{figure}[t!]
\centering
\includegraphics[width=0.49\textwidth]{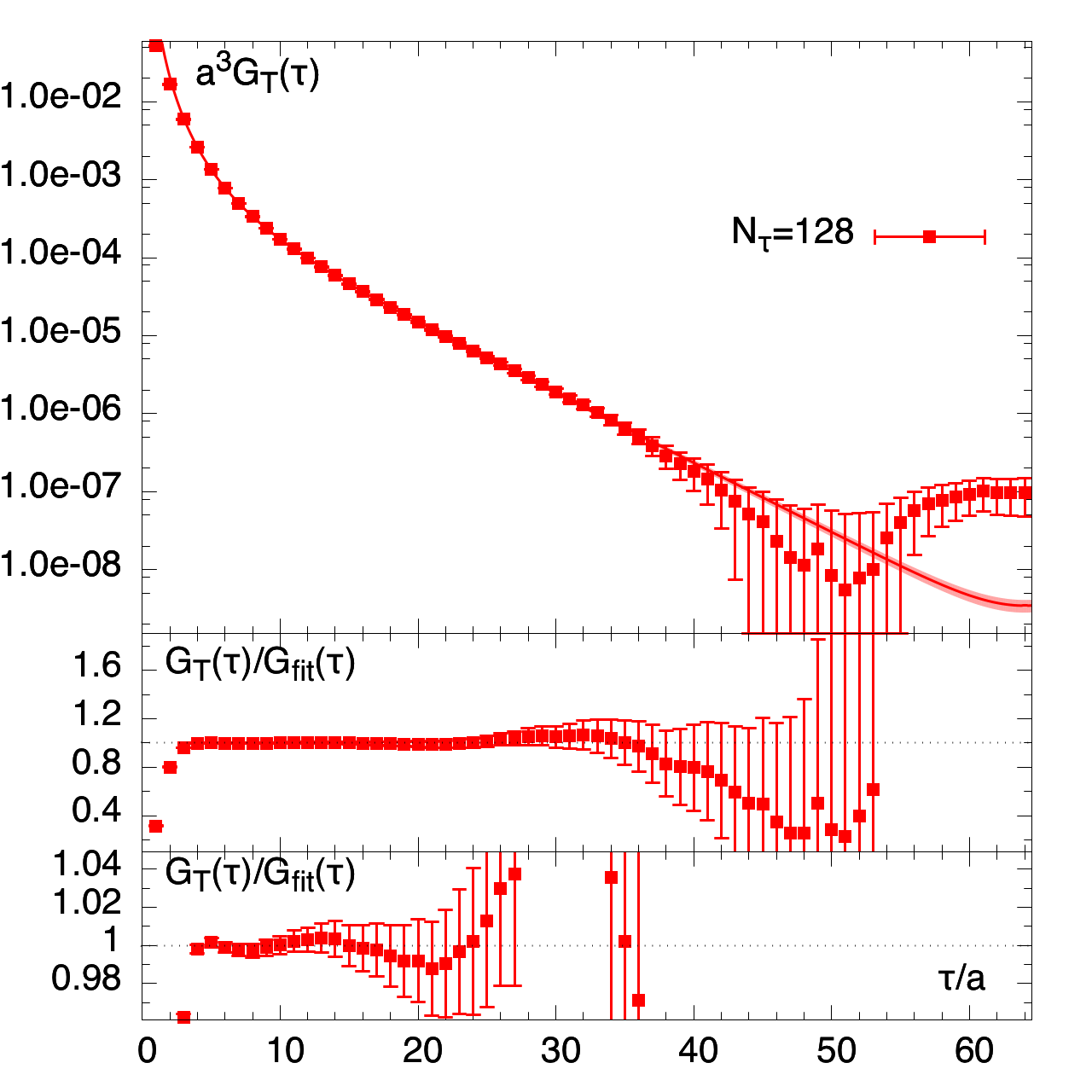}
\includegraphics[width=0.49\textwidth]{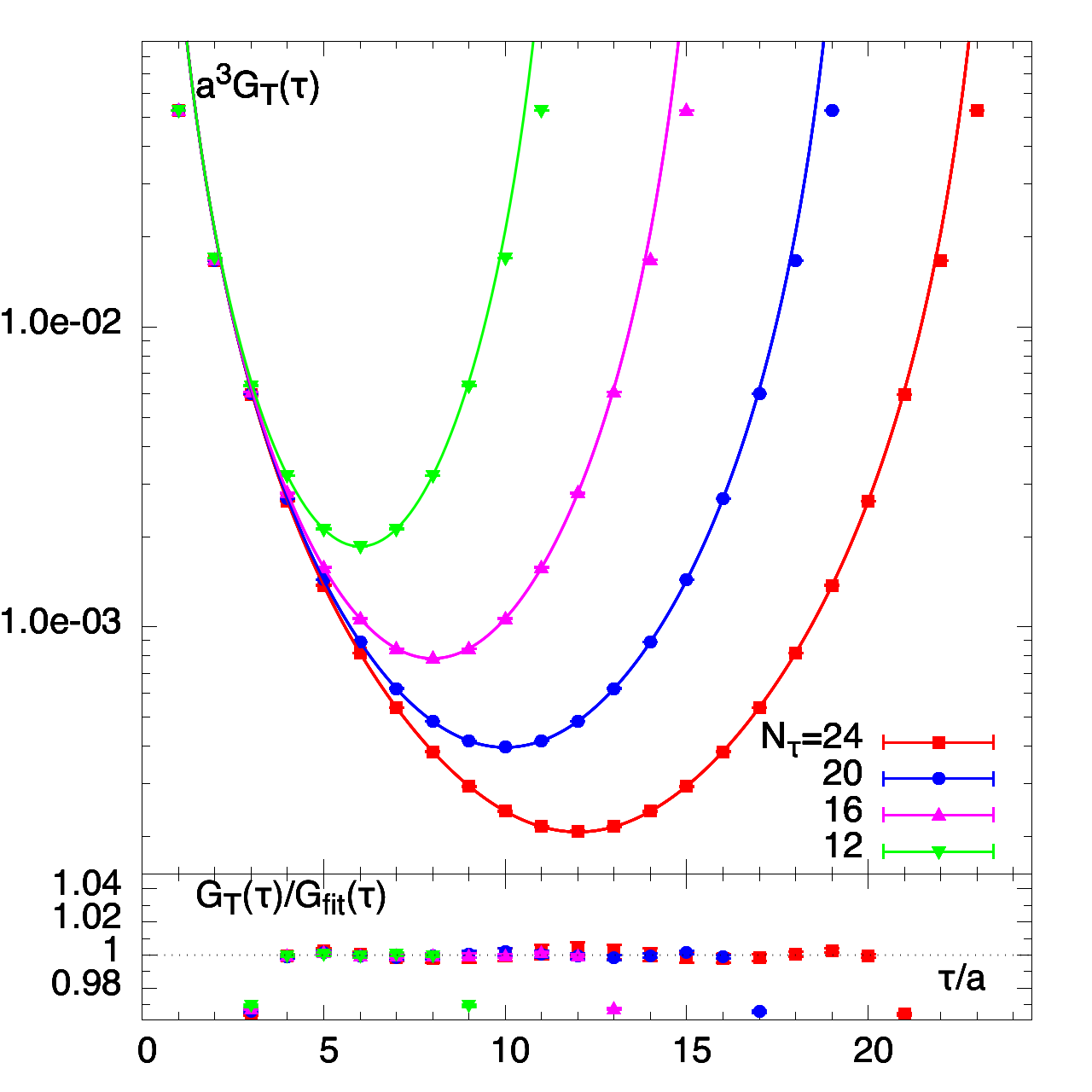}
\caption{Left: (Top) The vacuum ($N_\tau=128$) vector correlation
  function. The red lines denote the results computed by
  reconstructing the spectral functions with (Mod.\ 2c) from the lattice data. The
  middle panel shows the ratio of the data to the reconstructed
  result, the bottom panel shows a zoom of this ratio. We observe the
  lattice data is reproduced with a precision better than 2\% for the distance region
  $\tau/a\lesssim 20$. Right: (Top) The thermal vector
  correlators at $T/T_c=0.8, 1.0, 1.25$ and $1.67$, i.e. $N_\tau=24,
  20, 16$ and $12$. The lines denote the results 
  of the fits based on parametrizing the spectral functions (Mod.\ 2c). The
  bottom panel shows the ratio of the data to the fitted
  correlators.}
\label{fig:corr}
\end{figure}
%%%%%%%%%%%%%%%%%%%%%%%%%%%%%%%%%%%%%%%%%%%%%%%%%%%%%%%%%%%%%%%%

\subsubsection{Error estimation and results for the correlators}

The combined fit, as outlined above, is carried out on a large number of
bootstrap samples, each consisting of the average of $N=1000$ randomly
selected data sets, and uses a `frozen' statistical error obtained
from the fully sampled results. The parameters in the fit ans\"atze
are correlated, especially for the bound state and transport regions.
To give an error estimate for the parameters we use the method of quantiles as a consequence.  
As such, the central value parameters are  computed on the central value samples of the ensembles. 
Then we choose the parameter
sets corresponding to the central $68\%$ of the distribution of all solutions as our final results, i.e. we
assign asymmetric error bars corresponding to one
standard deviation. 
For all derived quantities, such as the correlators recalculated from the spectral functions, 
we show the standard bootstrapped results in the following.
Note, the quoted values for $\chi^2/$d.o.f are `uncorrelated' values and are collected in Tab.~\ref{tab:fitpar}.

\begin{figure}[t!]
\centering
\includegraphics[width=0.49\textwidth]{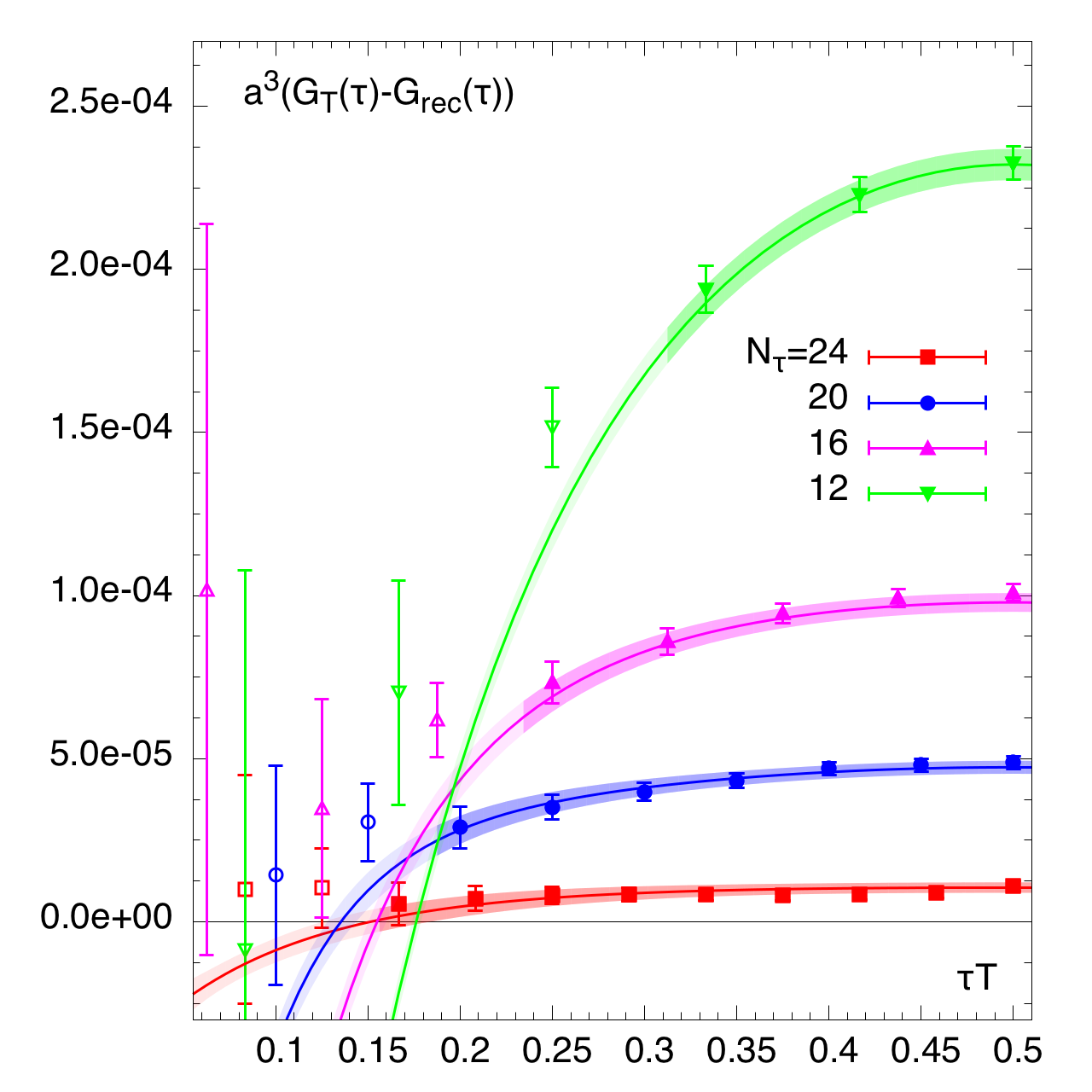}
\includegraphics[width=0.49\textwidth]{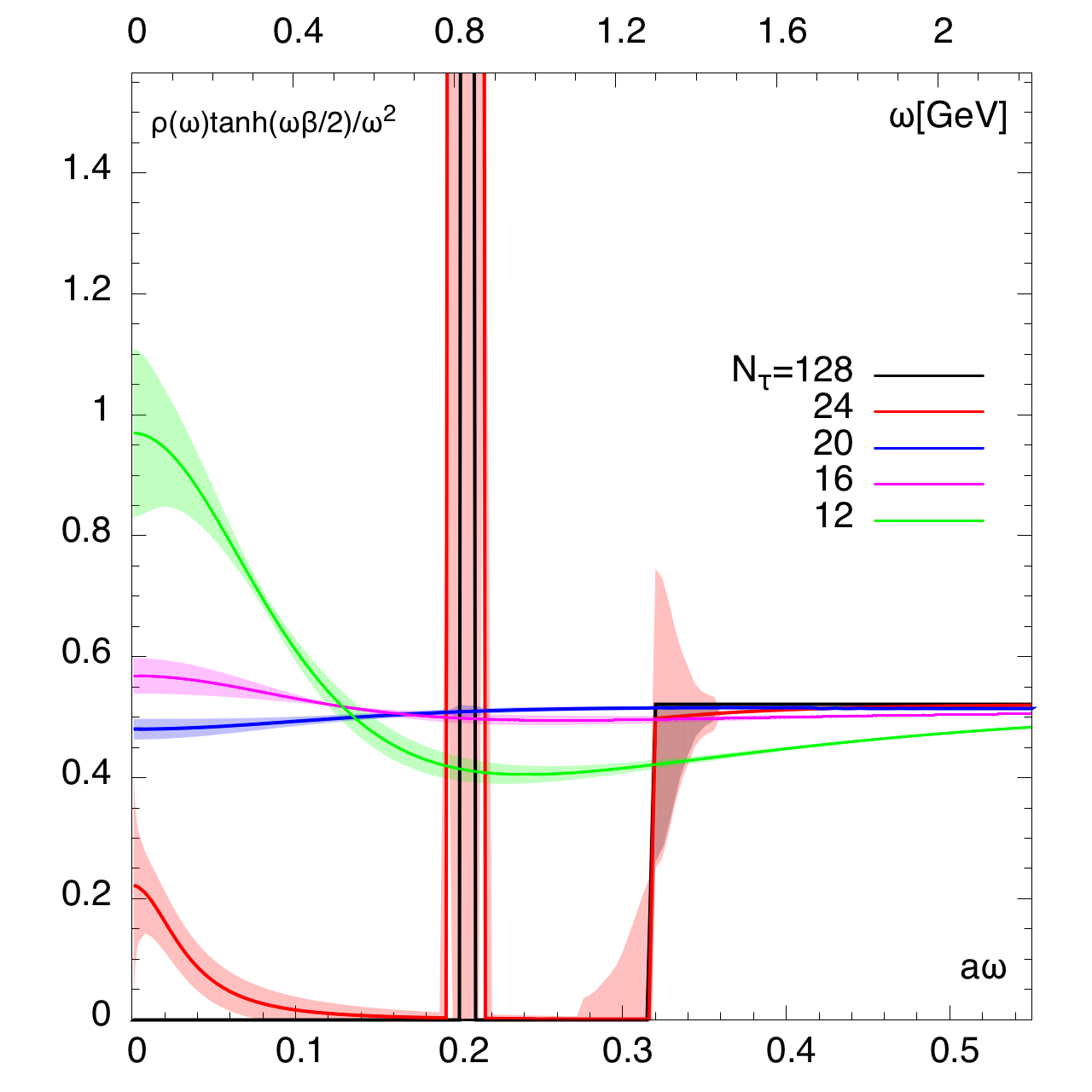}
\caption{Left: The difference of thermal and reconstructed correlators
  together with the results computed by reconstructing the spectral
  functions from the lattice data. The thermal vector correlators
  cover the temperature range $T/T_c=0.8, 1.0, 1.25$ and $1.67$,
  i.e. $N_\tau=24, 20, 16$ and $12$, where $T_c\simeq 203$MeV and
  $M_\pi\simeq270$MeV. The darker shaded region denotes the fit window
  used in the reconstruction (Mod.\ 2c). Right: The reconstructed spectral
  functions, rescaled by $\tanh(\omega/2T)/\omega^2$, in the vacuum (black) and thermal (colors)
  scenarios from (Mod. 2c).}
\label{fig:spf}
\end{figure}

The vacuum correlator is shown in Fig.~\ref{fig:corr} (left panel).  In the top panel of the figure we
show the correlator data and the corresponding fit. The
middle panel shows the ratio of the data to the reconstructed
correlators, the bottom panel shows a zoom of this ratio. We find that our
model describes the data accurately within $2\%$ for distances up to
$\tau/a\simeq 20$, i.e. $\simeq 1$fm. The corresponding spectral
function is displayed in \fig\ref{fig:spf} (right panel). 

Likewise, the resulting thermal correlators are shown in Fig.~\ref{fig:corr} (right panel).
In the top panel of this figure we show the correlator data and the
corresponding fit, while in the bottom panel we show the
ratio of the data and the fit. We observe the fitted model describes
the lattice data better than $1\%$ within the fit range.  
The thermal spectral functions are also shown in
Fig.~\ref{fig:spf} (right panel).

%%%%%%%%%%%%%%%%%%%%%%%%%%%%%%%%%%%%%%%%%%%%
%%%%%%%%%%%%%%%%%%%%%%%%%%%%%%%%%%%%%%%%%%%%

%%%%%%%%%%%%%%%%%%%%%%%%%%%%%%%%%%%%%%%%%%%%
\begin{figure}[t!]
%\centering
\includegraphics[width=0.329\textwidth]{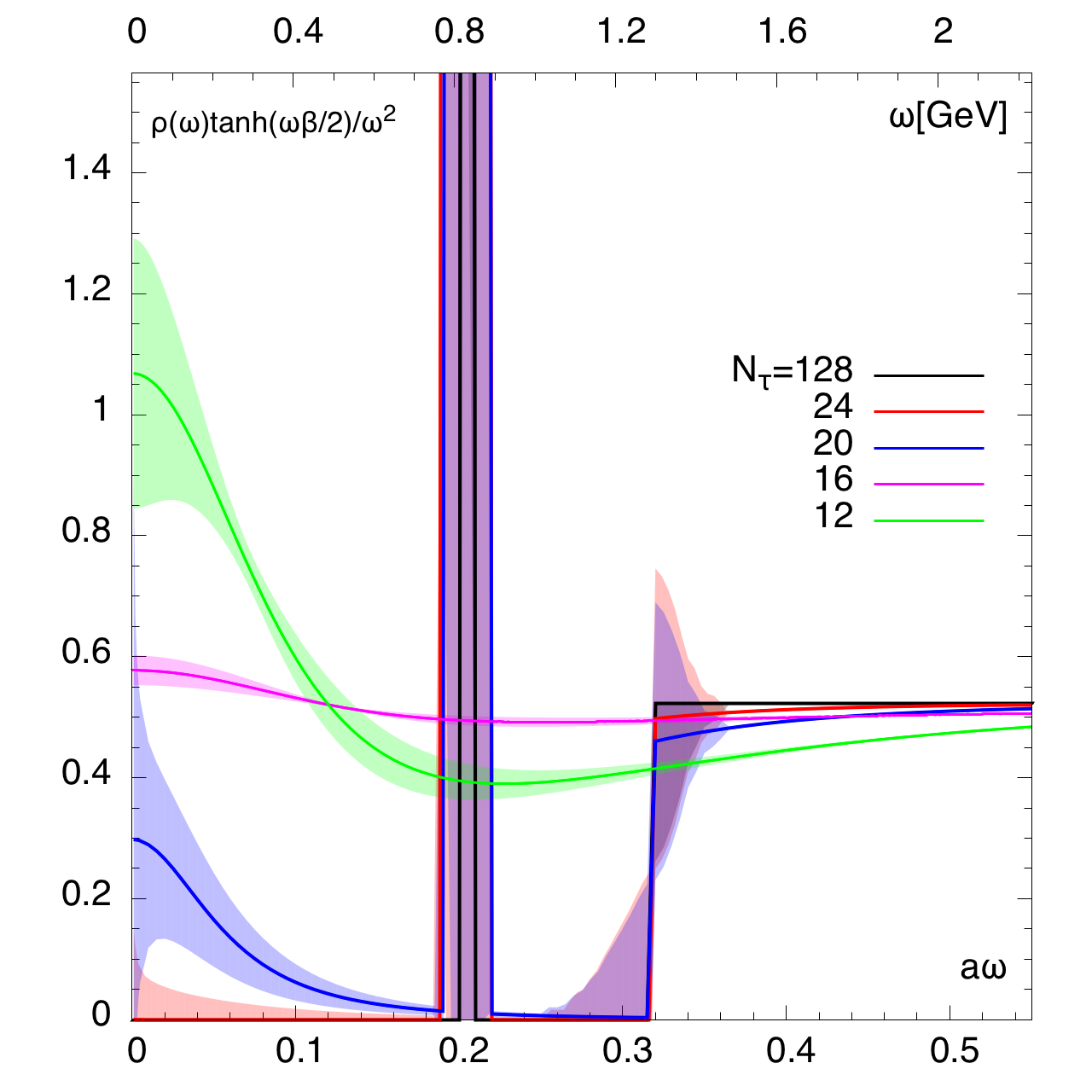}
\includegraphics[width=0.329\textwidth]{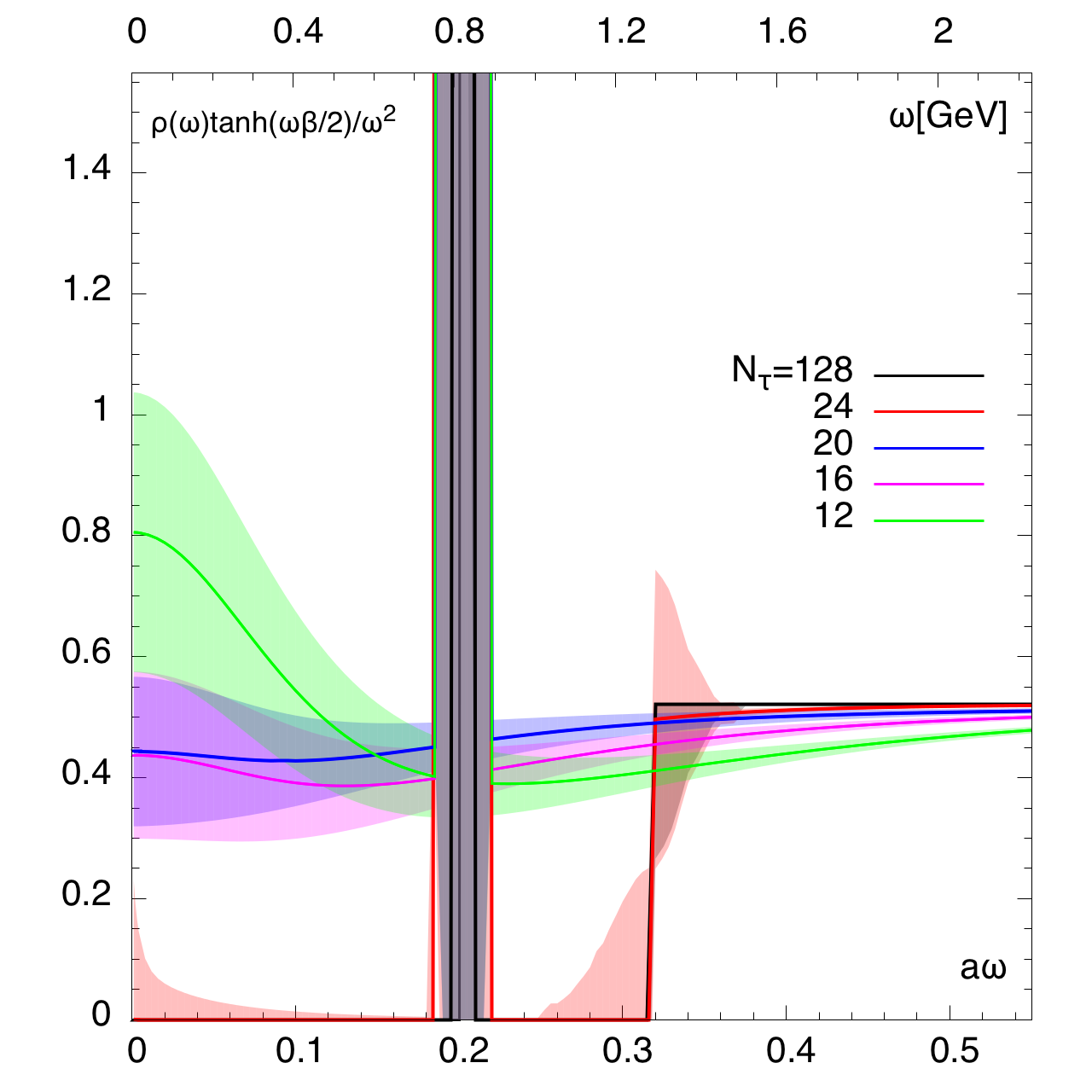}
\includegraphics[width=0.329\textwidth]{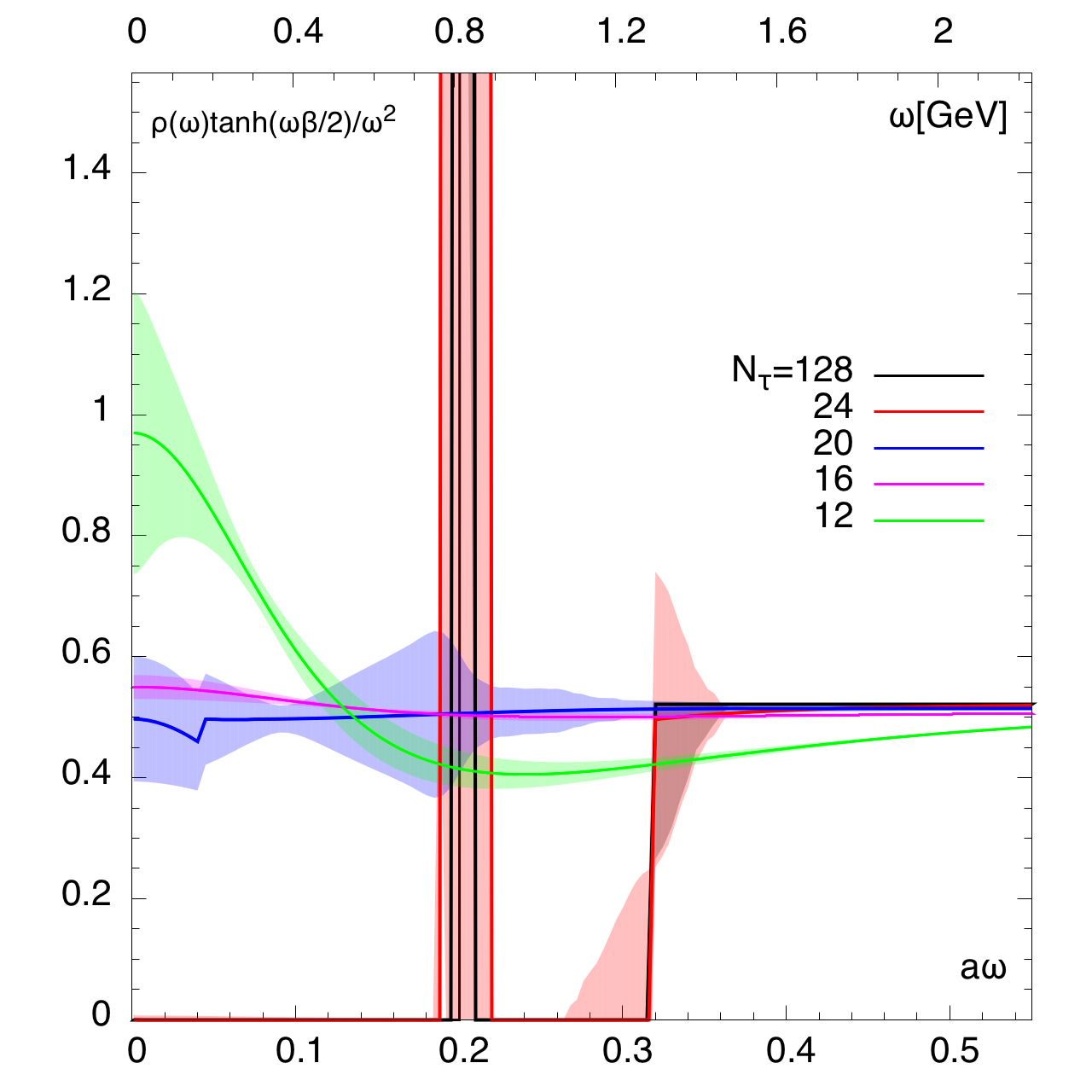}
\caption{The reconstructed spectral functions in the vacuum (black) and thermal (colors)
  scenarios from (Mod. 2a) in the left panel, (Mod. 2b) in the middle and from (Mod. 2d) in the right panel.}
\label{fig:spf2}
\end{figure}
%%%%%%%%%%%%%%%%%%%%%%%%%%%%%%%%%%%%%%%%%%%%

%%%%%%%%%%%%%%%%%%%%%%%%%%%%%%%%%%%%%%%%%%%%
\begin{table}[h!]
\centering
 % Give a unique label
% For LaTeX tables use
\begin{tabular}{llllll}
\hline\hline\noalign{\smallskip}
Par. [latt. units] & (Mod. 1) & (Mod. 2a)  & (Mod. 2b) & (Mod. 2c) & (Mod. 2d) \\
\noalign{\smallskip}\hline\noalign{\smallskip}
$m_V$ 				   & $0.205_{5}^{5}$ 	& $0.205_{5}^{4}$ & $0.205_{6}^{5}$ & $0.206_{4}^{3}$ & $0.205_{6}^{5}$  \\
\noalign{\smallskip}\hline\noalign{\smallskip}
$a_V$  					 & $0.00085_{9}^{8}$ & $0.00085_{8}^{7}$ & $0.00084_{10}^{8}$ & $0.00085_{6}^{5}$ & $0.00085_{9}^{8}$  \\
$a_{T,N_\tau=24}$  & $0.00086_{8}^{8}$ & $0.00086_{10}^{7}$ & $0.00086_{11}^{9}$ & $0.00086_{7}^{6}$ & $0.00086_{10}^{9}$  \\
$a_{T,N_\tau=20}$  & $0.00095_{9}^{8}$ & $0.00093_{9}^{8}$ & $0.00011_{5}^{4}$ & $2.8e^{-9}|_{2.5e^{-9}}^{1.2e^{-6}}$ & -  \\
$a_{T,N_\tau=16}$  & $0.00054_{5}^{5}$ 	& - & $0.00024_{4}^{8}$ & - & -  \\
$a_{T,N_\tau=12}$  & $0.00048_{11}^{12}$ & - & $0.00014_{7}^{9}$ & - & -  \\
\noalign{\smallskip}\hline\noalign{\smallskip}
$\Gamma_{T,N_\tau=24}$  & - & $1.0e^{-7}|_{1.0e^{-7}}^{5.0e^{-7}}$ & $1.0e^{-7}|_{1.0e^{-7}}^{5.0e^{-1}}$ & $0.03_{3}^{19}$ & $1.0e^{-7}|_{1.0e^{-7}}^{3.1e^{-7}}$  \\
$\Gamma_{T,N_\tau=20}$  & - & $0.062_{62}^{19}$ & $0.174_{44}^{170}$ & $0.269_{18}^{17}$ & $0.243_{20}^{20}$  \\
$\Gamma_{T,N_\tau=16}$  & - & $0.245_{25}^{52}$ & $0.168_{16}^{72}$ & $0.259_{33}^{24}$ & $0.284_{27}^{25}$  \\
$\Gamma_{T,N_\tau=12}$  & - & $0.109_{24}^{23}$ & $0.136_{17}^{95}$ & $0.126_{16}^{11}$ & $0.126_{24}^{26}$  \\
\noalign{\smallskip}\hline\noalign{\smallskip}
$\kappa_{\rm O,N_\tau=24}$  & $-0.00014_{25}^{22}$ & $-0.00014_{27}^{22}$ & $-0.00015_{3}^{7}$ & $-0.00014_{19}^{28}$ & $-0.00017_{24}^{23}$  \\
$\kappa_{\rm O,N_\tau=20}$  & $-0.00135_{21}^{20}$ & $-0.00155_{29}^{34}$ & - & - & - \\
$\kappa_{\rm O,N_\tau=16}$  & $-0.00281_{37}^{37}$ & - & - & -  & - \\
$\kappa_{\rm O,N_\tau=12}$  & $-0.00134_{45}^{50}$ & - & - & -  & -\\
\noalign{\smallskip}\hline\noalign{\smallskip}
$\kappa_0$  & $1.092_{3}^{2}$ & $1.093_{3}^{2}$ & $1.092_{4}^{3}$ & $1.092_{2}^{2}$ & $1.091_{3}^{3}$  \\
$\kappa_{1,N_\tau=20}$  & - & - & $-0.058_{38}^{13}$ & $-0.071_{4}^{4}$ & $-0.060_{11}^{25}$  \\
$\kappa_{1,N_\tau=16}$  & - & $-0.089_{26}^{9}$   & $-0.072_{29}^{17}$ & $-0.094_{8}^{5}$ & $-0.100_{13}^{10}$  \\
$\kappa_{1,N_\tau=12}$  & - & $-0.045_{11}^{11}$ & $-0.081_{43}^{7}$ & $-0.054_{8}^{12}$ & $-0.051_{11}^{11}$  \\
\noalign{\smallskip}\hline\noalign{\smallskip}
$\Omega_0$  & $0.319_{16}^{13}$ & $0.320_{15}^{11}$ & $0.317_{19}^{14}$ & $0.319_{11}^{9}$ & $0.318_{18}^{14}$  \\
$\Omega_{T,N_\tau=20}$  & - & - & - & - & $0.045_{45}^{180}$  \\
$\eta$  & - & $2.30_{12}^{17}$ & $2.01_{13}^{32}$ & $2.30_{9}^{14}$ & $2.36_{17}^{21}$  \\
\noalign{\smallskip}\hline\noalign{\smallskip}
%$\chi^2/d.o.f.$  & $0.56_{25}^{35}$ & $0.60_{23}^{40}$ & $0.54_{26}^{41}$ & $0.54_{24}^{40}$ & $0.51_{22}^{37}$  \\
$\chi^2/d.o.f.$  & $0.56$ & $0.60$ & $0.54$ & $0.54$ & $0.51$  \\
\noalign{\smallskip}\hline\hline
% \noalign{\smallskip}\hline
\end{tabular}
\caption{{The parameters in lattice units obtained by fitting the lattice vector meson correlators simultaneously and enforcing the sum rule. The upper and lower errors in the other columns indicate the top and bottom margins from the generally asymmetric error determined by the quantile method. As noted in the text the final results quoted originate from (Mod. 2c). We tested additional model setups and show only those explored fit ans\"atze that lead to an uncorrelated $\chi^2/{\rm d.o.f}\le 1.0$. }}
\label{tab:fitpar}      
\end{table}
%%%%%%%%%%%%%%%%%%%%%%%%%%%%%%%%%%%%%%%%%%%%

%%%%%%%%%%%%%%%%%%%%%%%%%%%%%%%%%%%%%%%%%%%%
\begin{table}[h!]
\centering
 % Give a unique label
% For LaTeX tables use
\begin{tabular}{llllll}
\hline\hline\noalign{\smallskip}
A$[a^2]$~~~~ & (Mod. 1)~~ & (Mod. 2a)~~  & (Mod. 2b) ~~& (Mod. 2c)~~ & (Mod. 2d)~~ \\
\noalign{\smallskip}\hline\noalign{\smallskip}
$A_{N_\tau=24}$  & $0.00023_{7}^{7}$ & $0.00023_{11}^{14}$ & $0.00022_{42}^{20}$ & $0.00032_{22}^{34}$ & $0.00021_{10}^{10}$  \\
$A_{N_\tau=20}$  & $0.00062_{12}^{12}$ & $0.00091_{22}^{23}$ & $0.00386_{77}^{285}$ & $0.00646_{28}^{31}$ & $0.00604_{277}^{80}$  \\
$A_{N_\tau=16}$  & $0.00095_{37}^{32}$ & $0.00883_{58}^{129}$ & $0.00460_{100}^{211}$ & $0.00918_{66}^{57}$ & $0.00977_{66}^{59}$  \\
$A_{N_\tau=12}$  & $0.00376_{87}^{76}$ & $0.00970_{66}^{66}$ & $0.00913_{193}^{276}$ & $0.01019_{50}^{33}$ & $0.01019_{68}^{70}$  \\
\noalign{\smallskip}\hline\hline
% \noalign{\smallskip}\hline
\end{tabular}
\caption{{The resulting height parameters of the transport contribution $A_{\rm T}[a^2]$ in lattice units from combined fitting the lattice vector meson correlators and enforcing the sum rule. As before, the upper and lower errors in the other columns indicate the top and bottom margins from the generally asymmetric error determined by the quantile method. }}
\label{tab:fitpar2}      
\end{table}
%%%%%%%%%%%%%%%%%%%%%%%%%%%%%%%%%%%%%%%%%%%% 

\subsubsection{Discussion of fit results\la{sec:fitres}}

We show the results from fitting model (Mod. 2c) to
$G(\tau,T)-G_{\rm rec}(\tau,0)$ in the left panel of
Fig.~\ref{fig:spf}. Here, the shaded bands denote the results from the
spectral function reconstruction using the ansatz approach.  The data
used in the fits is shown in a darker shade and by filled symbols,
while the lighter shaded regions and open symbols denote the
region outside of the fit windows.  For all
temperatures we observe good agreement between the data and the
results from the fitted spectral functions.

In the right panel of Fig.~\ref{fig:spf} and in Fig.~\ref{fig:spf2} we show the reconstructed
spectral functions for (Mod. 2a-d), rescaled by
$\rho(\omega)\rightarrow\rho(\omega)\tanh(\omega\beta/2)/\omega^2$. This
rescaling renders the displayed functions finite both at low and large
frequencies.

Focusing on low frequencies, i.e. the transport region of the spectral
functions in Fig.~\ref{fig:spf} (right) and Fig.~\ref{fig:spf2}, we observe a significant contribution of
spectral weight in the transport region for all tested variants above $T\geq T_c$. The intercepts of the spectral functions for the $N_\tau=20$ ensemble overlap well with each other, where the most vacuum-like variant (Mod. 2b) in Fig.~\ref{fig:spf2}(left) appears to be poorly constrained in this region, due to its large uncertainties.
In the confined phase ($N_\tau=24$), we observe a tendency for the suppression of the transport peak and a small change of the bound state peak compared to the vacuum case.
Below $\omega=1\,$GeV, we observe a "filling-up" of spectral weight, leading to almost flat spectral functions for $N_\tau=20$ and $N_\tau=16$. 
The almost constant behavior of the function $\rho(\omega)\tanh(\omega\beta/2)/\omega^2$ is reminiscent of the spectral function 
of the R-charge current in the strongly coupled  ${\cal N}=4$ super Yang-Mills~\cite{Myers:2007we};
such a spectral function describes a medium with no quasiparticles.
At the highest temperature ($N_\tau=12$ ensemble) we see an excess of spectral weight around the origin as compared to the flat behavior, 
suggesting that a transport peak is beginning to emerge. At even higher temperatures, 
the transport peak is expected to become gradually narrower in units of $\omega/T$.

The parameters determining the spectral functions are given in Tab.~\ref{tab:fitpar}, whereby the upper and lower numbers indicate the errors obtained using the quantile method.
The amplitudes of the bound state peaks between the vacuum and $N_\tau=24$ ensemble are very close to each other, thereby indicating no clear sign of thermal modification. For (Mod. 1) and (Mod. 2a), where a bound state contribution is explicitly included, we observe an increase of the peak amplitude at $T=T_c$, i.e. for the $N_\tau=20$ ensemble. On the other hand, models based on a more thermal behavior show a strong suppression of the peak amplitude. The spectral weight lost around the $\rho$ mass is seemingly compensated by a modification of $\kappa_0$ via $\kappa_{1}$. The width of the transport peaks $\Gamma_{\rm T}$ are consistent in the last two models (Mod. 2c and Mod. 2d) in the deconfined phase. In addition we observe a trend from broader to narrower widths as the temperature increases. At $T_c$, the `thermal' models (2b, 2c, 2d) agree within errors, while the `vacuum' model (2a) exhibits larger uncertainties. The parameters in the confined phase show large uncertainties for all models.

The height parameters of the transport peak $A_{\rm T}$, derived using the sum rule, are listed in Tab.~\ref{tab:fitpar2}. The values using (Mod. 1) are consistently lower in the deconfined phase compared to the variants of (Mod. 2). At the same time they are compatible in the confined phase and at $T_c$ when a vacuum-like variant is used. This indicates a very narrow and suppressed transport contribution at $N_\tau=24$. 
For all variants of (Mod. 2) the values of $A_{\rm T}$ show consistent trends and agree at a qualitative level. As such we observe an increase of $A_{\rm T}$ with temperature for all models tested. 

Taken together, the suppression of a bound state peak in favor of a broad spectral distribution in (Mod. 2c) and the consistency between the results of (Mod. 2b)-(Mod. 2d) hint at a rapid dissociation of the narrow bound state as the system crosses over into the deconfined phase. At the same time we see large changes in the electrical conductivity, mirrored by
$A_{\rm T}$ at $T=T_c$, where the width of the transport peak acquires a large value that then decreases with temperature.

Although our results are not yet decisive enough to present a
definite, ab initio and precision determination of the dissociation of
the $\rho$ meson, this is, to our knowledge, the first lattice
calculation resolving the spectral weight of a light hadron as it crosses the deconfinement phase transition.

%%%%%%%%%%%%%%%%%%%%%%%%%%%%%%%%%%%%%%%%%%%%%%%%%%%%%%%%%%%%%%%%
%%%%%%%%%%%%%%%%%%%%%%%%%%%%%%%%%%%%%%%%%%%%%%%%%%%%%%%%%%%%%%%%

\subsection{The Backus-Gilbert method\la{sec:BG}}

Up to this point the results depend on the chosen model for the true
spectral function. Although we believe that our assumptions are
plausible, we present in the following a method to
constrain the spectral function locally in the variable $\omega$ using
only the information contained in the lattice data.  This can be
achieved by means of the Backus-Gilbert method (BGM)
\cite{Backus:1968bg}. The method represents a paradigm change from the
usual goal of reconstructing the entire spectral function from the
lattice correlation function data. We applied this method to the study of pion dispersion relations in \cite{Brandt:2015sxa} and now extend it to the vector spectral functions.

First, define a rescaled kernel,
\ba
K(\tau,\omega) &=& f(\omega/T)\,\frac{\cosh[\omega(\beta/2-\tau)]}{\sinh{[\omega\beta/2]}},
\ea
where $f(x)\stackrel{x\to0}{\propto} x$ and $f(x)>0$ for $x>0$. Appropriate choices for $f(x)$ will be discussed later.
Now consider a `resolution function' $\delta(\omega,\omega')$, such that the `filtered spectral function'
\be\la{eq:whatrho}
\widehat\rho(\omega) \equiv f(\omega/T)\int_0^\infty d\omega'\, \delta(\omega,\omega') \,\frac{\rho(\omega',T)}{f(\omega'/T)}
\ee
provides an averaged value (in a sense made more precise below) of $\rho$ around frequency $\omega$.
For a given $\omega$, the goal is for $\delta(\omega,\omega')$, viewed as a function of $\omega'$,
to be  a narrowly concentrated function around  $\omega$.
Restricting ourselves to methods linear in the lattice data, the filtered spectral function
is necessarily given by a linear combination of the form
\be\la{eq:rhohatG}
\widehat\rho(\omega)  = f(\omega/T) \sum_{i=1}^n g_i(\omega) \,G(\tau_i),
\ee
where $\tau_i$ are the Euclidean times, for which the correlator has been calculated.
The coefficients $g_i(\omega)$ are chosen to realize the stated goal.
Inserting the Euclidean correlator into Eq.~(\ref{eq:rhohatG}) in terms of $\rho(\omega',T)$,
we conclude that the resolution function is given by
\be\la{eq:resfct}
\delta(\omega,\omega') = \sum_{i=1}^n g_i(\omega) \,K(\tau_i,\omega')~~.
\ee
For a given $\omega$, it is thus completely specified by the $n$ coefficients $g_i(\omega)$.
One possible recipe to construct the $g_i(\omega)$ is provided by 
the method of Backus and Gilbert \cite{Backus:1968bg}, which minimizes the `width'
\be\la{eq:Gw}
\Gamma_\omega\equiv \int_0^\infty d\omega' \,(\omega-\omega')^2 \,\delta(\omega,\omega')^2 
\ee
of the resolution function for a normalized area
\be\la{eq:Ad1}
\int_0^\infty d\omega' \delta(\omega,\omega') = 1.
\ee
The latter condition implies that, if $ \rho(\omega',T)/f(\omega'/T)$ is independent of $\omega'$ for 
$|\omega'-\omega|\lesssim \Gamma_\omega$, the filtered spectral function
$\widehat\rho(\omega)$ is an estimator for the value of
$\rho(\omega,T)$ at the point $\omega$. The solution is (see for instance~\cite{Press:2007zz})
\be
g_i(\omega) = \frac{(W^{-1})_{ij}R_j}{R_k (W^{-1})_{kl} R_l}, 
\ee
with 
\be
 W_{ij}(\omega)=\int_0^\infty d\omega'\, K(\tau_i,\omega') K(\tau_j,\omega') (\omega-\omega')^2~~~
\textrm{and}~~~ R_i \equiv \int_0^\infty d\omega\, K(\tau_i,\omega)~~.
\ee
In practice, the matrix $W$ rapidly becomes ill-conditioned with increasing $n$,
the coefficients $g_i(\omega)$ become large and alternating in sign, which is a manifestation
of the ill-posed nature of the inverse problem.
It is necessary to regularize the procedure by replacing $W$ with 
\be
W^{\rm reg.}_{ij}(\omega) = \lambda \,W_{ij}(\omega) + (1-\lambda) \,{\rm Cov}_{ij}\,[G]~~,\qquad 
0\leq \lambda\leq 1,
\ee
where ${\rm Cov}[G]$ is the covariance matrix of the lattice correlator $G(\tau_i,T)$.
The Backus-Gilbert method can then be thought of as minimizing the width $\Gamma_\omega$
under the condition that the statistical error on $\widehat\rho(\omega,T)$ assumes a preset value.
Increasing $\lambda$ slowly reduces $\Gamma_\omega$ at the cost of increasing
the statistical errors.
The only input into the construction of the resolution function 
is thus the covariance matrix, the choice of $f(x)$ and the 
value of the regularization parameter $\lambda$.  
Independent of the choices made, it is clear that quoting  $\widehat\rho(\omega,T)$ at $n$ 
values of $\omega$ -- typically spaced by separations of order $\Gamma_\omega$ -- is equivalent
information to the original lattice data; one has simply deconvoluted the information
as local in $\omega$ as the quality of the lattice data allows.
We note that the resolution function has already been used in~\cite{Meyer:2007dy} as a way 
to provide a rigorous meaning to a linear reconstruction method for the spectral function.

We now discuss the choice of the reweighting function $f(x)$.
It is desirable to choose it in such a way, that the reconstructed function, $\rho(\omega,T)/f(\omega/T)$, does not show
a global trend. For the vector channel considered here,
we note that
$(\rho_{ii}(\omega,T) \frac{T^2}{\omega^2}\tanh\frac{\omega}{2T} )$
is expected to be finite in both limits, $\omega\to 0$ and $\omega\to
\infty$.  We therefore choose 
\be\la{eq:Ktw1}
f(x) = \frac{x^2}{\tanh(x/2)}.
\ee
We will also consider the vacuum-subtracted correlation functions,
$G(\tau,T)-G_{\rm{rec}}(\tau,0)$.  Since we expect
$\Delta\rho(\omega)\stackrel{\omega\to\infty}{\propto} 1/\omega^2$ from the operator-product
expansion, we choose 
\be\la{eq:Ktw2}
f^\Delta(x) = \tanh(x/2)
\ee
as reweighting function.  Various reweighting functions were already
introduced in
\cite{Aarts:2007wj,Meyer:2007ic,Engels:2009tv,Ding:2012sp,Brandt:2015sxa} to remove
the divergence of the kernel at the origin.  Here, the reweighting
function plays the additional role of rescaling the spectral function
to remove its overall growth at large $\omega$.

Finally, we remark that many other choices than Eq.~(\ref{eq:Gw}) are
possible as a measure of the width of $\delta(\omega,\omega')$.  The
most useful choice depends on the channel under consideration as well
as what question precisely one is addressing.

%%%%%%%%%%%%%%%%%%%%%%%%%%%%%%%%%%%%%%%%%%%%%%%%%%%%%%%%%%%%%%%%
\begin{figure}[t!]
\centering
\includegraphics[width=0.49\textwidth]{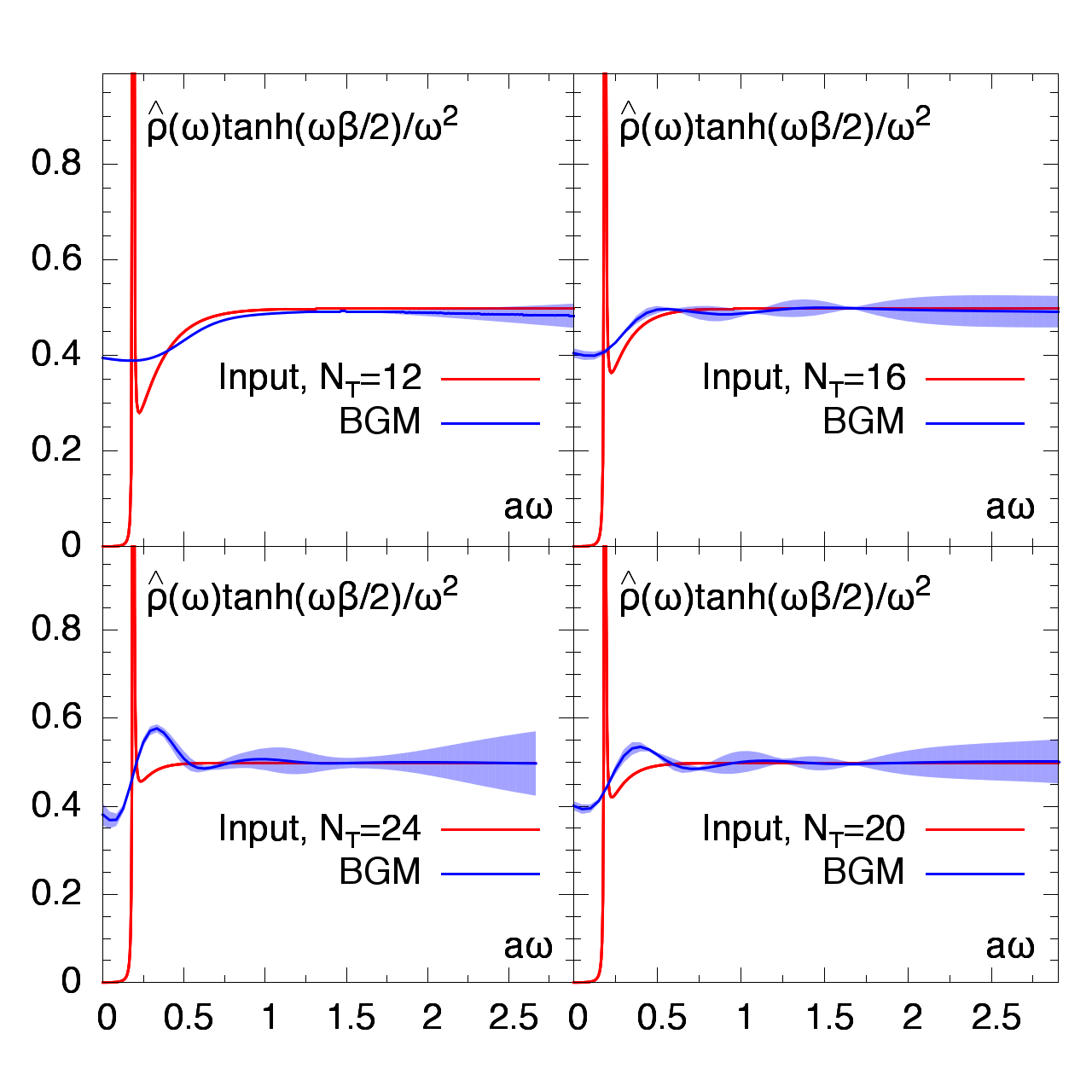}
\includegraphics[width=0.49\textwidth]{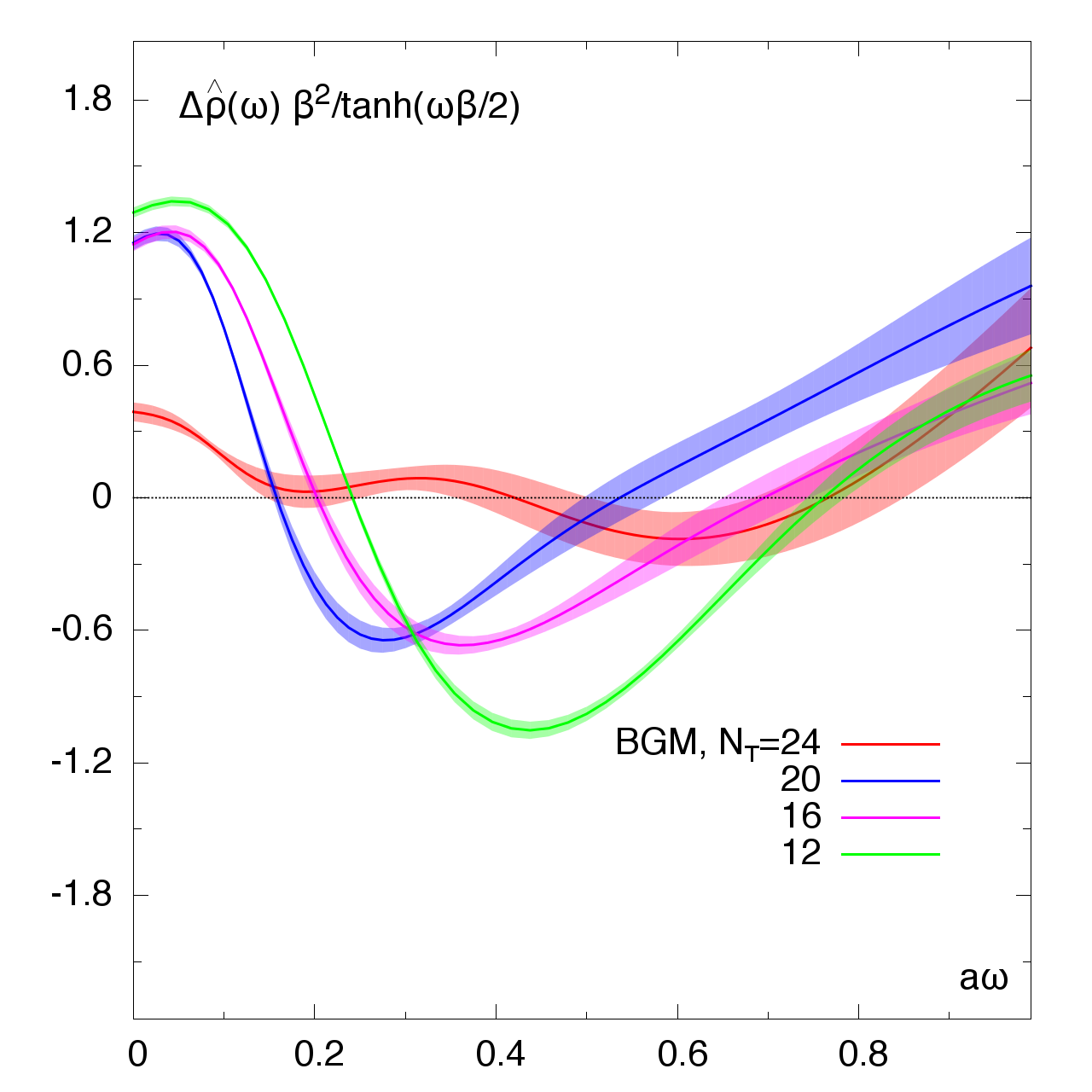}
\caption{Left: Mock data test of the Backus-Gilbert method, reconstructing $\widehat\rho(\omega)$ from a known spectral function (red) and 
the covariance matrix supplied by the lattice data.
Right: The convoluted spectral functions for the difference
  $\Delta\widehat\rho(\omega)$ from reconstructing
  $G(\tau,T)-G_{\rm{rec}}(\tau,0)$ using the BGM. The narrower
  resolution function and favorable cancellation of the large
  frequency regions enable more details to be reconstructed.}
\label{fig:bgm}
\end{figure}
%%%%%%%%%%%%%%%%%%%%%%%%%%%%%%%%%%%%%%%%%%%%%%%%%%%%%%%%%%%%%%%%

\subsubsection{Applying the BGM: tests and results}

We found that a value of the
regularization parameter $\lambda=0.002$ was a reasonable choice for
all ensembles, although larger values up to $\lambda=0.9$ lead to
equally good results for the high temperature lattices $N_\tau=12$ and
16. The regularization is crucial for the lower-temperature
lattices, when a large number of points enter $W_{ij}(\omega)$.

To illustrate the precision that can be achieved, we show
the resolution function $ T\delta(\omega,\omega')$ computed on
the $N_\tau=24$ ensemble, as a function of $\omega'/T$, for the four
values $\omega/T=0$, 4, 8 and 16 in Fig.~\ref{fig:phencomp} (left
panel). As input we choose the points $\tau_i = (3,...,12)$ in
$G(\tau)$.  We observe that the width of the resolution function increases
with increasing frequency $\omega$, from roughly $\Gamma_\omega\sim
5T$ at $\omega/T=0$ to $\Gamma_\omega\sim 15T$ at $\omega/T=8$. Since
the resolution function only depends on the kernel and the precision
of the lattice data, this sets limits on the frequency
resolution that can be reached in our lattice calculation. 
The relatively poor resolution is partly due to the rise $\sim\omega^2$
of the spectral function of the conserved vector current.

Next we perform a mock data test to assess the relation between $\widehat\rho(\omega,T)$
and $\rho(\omega,T)$ in the vector case. To this end we
replace the average values of the lattice correlator by pseudodata
derived from a known spectral function, which we chose to be the
fitted vacuum result, and keep the lattice covariance matrix.  As
before the points chosen as input to the method are
$\tau_i=(3,\dots,N_\tau/2)$.  The resulting filtered spectral
functions, $\widehat{\rho}(\omega)$, are shown in
Fig.~\ref{fig:bgm} (left panel) for all four available lattice
ensembles. With increasing temperature we observe the sharp features
of the input mock spectral function to be washed out by the filtering
effect of the resolution function.

In the last step, we apply the Backus-Gilbert method to actual lattice data,
both for the correlator $G(\tau,T)$ itself and for the difference
$G(\tau,T)-G_{\rm{rec}}(\tau,0)$.  The results of the latter case are
displayed in red in Fig.~\ref{fig:bgm} (right panel), while the former are given in Fig.~\ref{fig:phencomp} (left). The narrower
resolution function and the cancellation of the large frequency
behavior in $\Delta\rho(\omega,T)$ allow more structure to be
extracted from the lattice data. See the right panel of
\fig\ref{fig:phencomp} for the corresponding resolution function at
$T=169\,$MeV.  As the temperature is increased we observe the
emergence of a negative dip around the vector meson mass, along with
an enhancement in the transport region. This suggests the
dissociation of the $\rho$ meson and the formation of a transport
peak. This dip forms already at $N_\tau=20$, corroborating to the conclusion of section \ref{sec:fitres}, that there is a rapid dissociation of the bound state around $T_c$ accompanied by a rapid build-up of the transport contribution.

%%%%%%%%%%%%%%%%%%%%%%%%%%%%%%%%%%%%%%%%%%%%%%%%%%%%%%%%%%%%%%%%

\section{Comparisons with model predictions\la{sec:compa}}

\subsection{Comparing lattice results to the HRG model\la{sec:vsq}}
As discussed in section \ref{sec:theory}, the hadron resonance gas
(HRG) model makes definite predictions for the the static
susceptibility $\chi_s$ and the mean-square-velocity of the hadronic
quasiparticles in the low-temperature phase of QCD. Here we compare
these predictions to the results obtained from our lattice simulations.

On the lattice, the static susceptibility is
computed straightforwardly using Eq.\ (\ref{eq:chis}).
The mean-square velocity is more difficult to estimate, since it must be extracted 
from the spectral function and assumes the presence of a well-defined transport peak.
To construct an estimator for the mean-square velocity, we use the BG filtered spectral function $\widehat{\rho}(\omega,T)$
to write
\be\la{eq:v2est}
\<v^2\>_{\rm eff} 
= \frac{1}{2\pi\chi_s\; T\,\delta(0,0)} \; \frac{ \Delta\widehat{\rho}(\omega,T)}{\tanh(\omega \beta/2) }\Big|_{\omega=0}~,
\ee
which is based on Eq.~(\ref{eq:IntLbda}). It relies on the narrowness
of the transport peak compared to the width of the resolution function
$\delta(0,\omega)$.  We focus the following discussion on the
model independent results obtained using the BGM and present the static susceptibility, the value of
$T\cdot\delta(0,0)$ for the resolution function and the mean-square
velocities in Tab.~\ref{tab:chisv^2}.

%%%%%%%%%%%%%%%%%%%%%%%%%%%%%%%%%%%%%%%%
\begin{table}[t!]
\begin{tabular}{llcl@{~~~}}
\hline\hline\noalign{\smallskip}
$T[{\rm MeV}]$    &   $\chi_s/T^2$  &  $T\delta(\omega_0=0,\omega=0)$ 
& $\sqrt{\<v^2\>_{\rm eff}}$  \\
\noalign{\smallskip}\hline\noalign{\smallskip}
169   &   0.42718(31)   & 0.36588 &   0.629(34)  \\
203   &   0.72120(29)   & 0.37275 &   0.826(12)  \\
254   &   0.86771(34)   & 0.37484 &   0.749(10)  \\
338   &   0.93949(25)   & 0.37250 &   0.7661(68) \\
\noalign{\smallskip}\hline\hline
\end{tabular}
\caption{The static susceptibility and the effective root-mean-square velocities,
obtained on the four thermal lattice ensembles using the BGM filtered
thermal spectral functions.
The resolution function is given in the third column.}
\label{tab:chisv^2}
\end{table}
%%%%%%%%%%%%%%%%%%%%%%%%%%%%%%%%%%%%%%%%

To obtain an HRG prediction at the physical $(u,d)$ quark masses, we
would sum up the contributions to $\chi_s$ and $\chi_s\<v^2\>$ of the
mesons and baryons listed in the Particle Data Group
\cite{Nakamura:2010zzi} up to a mass of about 2GeV.  Since the quark masses
in our lattice simulations are larger than in nature, we attempt to
correct for the bulk of this effect by setting the pion, $\rho$ and
nucleon masses to the values found in our lattice ensembles. For the
other mesons and baryons, we have added a quark-model inspired shift
to the PDG masses equal to $m^{\rm latt}_\rho-m_\rho^{\rm phys}$ and
$m^{\rm latt}_N-m_N^{\rm phys}$ respectively.  Since we perform
two flavor simulations, hadrons with non-vanishing strangeness or charm
are not included.  We thus obtain the following HRG estimates for the
two-flavor theory with a zero-temperature pion mass of $\simeq270$MeV,
\be
{\rm HRG:}\qquad 
\chi_s/T^2 = \left\{\begin{array}{c@{~~~~}c}  
              0.49 &   T=169{\rm MeV}   \\  0.93 & T=203{\rm MeV}
              \end{array} \right.,
\qquad 
\sqrt{\<v^2\>} = \left\{\begin{array}{c@{~~~~}c}
              0.74 &   T=169{\rm MeV}   \\  0.70 & T=203{\rm MeV}
              \end{array} \right.~.
\la{eq:HRGv2}
\ee
The temperature evolution of the mean-square velocity originates from 
two competing effects: As the temperature rises, heavier hadrons have a 
chance to appear in the medium with a small velocity. On the other hand, 
lighter hadrons become gradually more relativistic. Thus a mean-square velocity which decreases
as a function of temperature signals the approach to the Hagedorn regime where 
the appearance of new hadron species dominates. In our model, $\<v^2\>$ has a maximum 
around $T=130\,$MeV.

Comparing the
static susceptibilities in Eq.~(\ref{eq:HRGv2}) and in
Tab.~\ref{tab:chisv^2}, we find a reasonably good agreement at
$T=169$MeV, but a clear overestimate of the HRG prediction at
$T=203$MeV.  The HRG value is in fact already close to the
Stefan-Boltzmann limit value of $\chi_s/T^2 = 1$.  

Looking at the mean-square velocities, we start with
the ensemble at $T=169$MeV.  The width of
$\delta(0,\omega)$ is approximately $\frac{\pi}{2} T$ in our study. In this
regime, the quantity $\<v^2\>_{\rm eff}$ is a good estimator of the
mean-square velocity of quasiparticles, if (a) the transport peak at
the origin in $\rho_{ii}(\omega,T)$ exists and is narrow in comparison
with the thermal scale $\omega\sim \pi T$ and (b) the next significant
contribution to the subtracted spectral function
$\Delta\rho(\omega,T)$ lies well beyond the width of
$\delta(0,\omega)$. As for (a), kinetic theory estimates\footnote{E.g. $\Gamma=\sigma n v$,
with $\sigma$ a typical hadronic cross-section and $n$, $v$ the density and the velocity 
of quasiparticles. Also, the thermal width of the pion quasiparticle  calculated 
in chiral perturbation theory~\cite{Schenk:1993ru} is quite small.} of the width
of the transport peak based on a hadronic gas picture clearly
indicate that the width is much smaller than $T$. 
Assuming this to be true at the lowest temperature $T=169$MeV,
we observe that the lattice estimate of $\sqrt{\<v^2\>}\approx 0.63$
is somewhat smaller than the HRG value 0.74. 
This observation is interesting, since we found on the same lattice
ensemble in our recent study~\cite{Brandt:2015sxa} that the pion quasiparticle is
lighter than the zero-temperature pion mass and has a less steep
dispersion relation, $\omega_{\vec p}^2 = {m_{\rm quasip}^2+u^2\vec p^2}$
with $u=0.74(1)$ and $m_{\rm quasip}=223(4)$MeV.
An alternative to the HRG model is then to only include the pion contribution, 
however taking into account the modified dispersion relation at low momenta,
\ba % \la{eq:chi_pion}
\chi_s &=& 4\beta \int_{|\vec p|<\Lambda_p} \frac{d^3\vec p}{(2\pi)^3} \; f^{_B}(\omega_{\vec p}) (1+f^{_B}(\omega_{\vec p}))
+ 4\beta \int_{|\vec p|>\Lambda_p} \frac{d^3\vec p}{(2\pi)^3} \; f^{_B}_{\vec p} (1+f^{_B}_{\vec p})~~,
\\ % \la{eq:chi_pion2}
\chi_s \<v^2\> &=& 4\beta \int_{|\vec p|<\Lambda_p} \frac{d^3\vec p}{(2\pi)^3} \; f^{_B}(\omega_{\vec p}) (1+f^{_B}(\omega_{\vec p}))\, 
v_g(\vec p)^2 
+ 4\beta \int_{|\vec p|>\Lambda_p} \frac{d^3\vec p}{(2\pi)^3} \; f^{_B}_{\vec p} (1+f^{_B}_{\vec p})\, \frac{\vec p^2}{E_{\vec p}^2}~~,
\ea
where $f^{_B}(E) = (e^{\beta E}-1)^{-1}$ 
and $v_g(\vec p) = \frac{d\omega_{\vec p}}{d|\vec p|}$ is the group velocity of the pion.
In this model, the contribution of the other hadrons are at least partly taken into account indirectly, since it is 
the collisions of the pions among themselves and with other hadrons which give rise to the 
modified pion dispersion relation. With a choice of $\Lambda_p=500\,$MeV, which is about the momentum scale at which
the predictions of the thermal chiral effective theory were seen to break down in~\cite{Brandt:2015sxa},
one obtains $\chi_s/T^2 = 0.43$ and $\sqrt{\<v^2\>}=0.68$, in better agreement with the lattice data than the HRG model.

There are however alternative explanations for the difference between the HRG prediction and the lattice result
for $\<v^2\>$. One explanation could be that our lattice value for $\<v^2\>$ is somewhat underestimated.
Indeed, this could happen if assumption (b) stated in the previous paragraph were imperfectly realized
and if the spectral weight of the $\rho$ meson was somewhat lower than at $T=0$.
After all we know from the sum rule in Eq.~(\ref{eq:sr}) that some negative spectral weight in $\Delta\rho(\omega,T)$ must
appear to compensate the transport peak. 
A further possible explanation that ought to be studied is the influence of the finite
volume in our simulation. Roughly, the average number of hadrons in
our $(3{\rm fm})^3$ box is of the order of seven, based on the HRG model,
which could invalidate a kinetic theory description.

At $T=203$MeV, we find using the lattice correlators
that the effective rms velocity has risen significantly from its
value at $T=169$MeV, while the HRG model predicts a slight decrease.
The respective values are 0.83 and 0.70. It is more difficult to explain
an overestimate from systematic effects affecting $\sqrt{\<v^2\>_{\rm eff}}$.
Considering the static susceptibility, the rms velocity
and our finding in section \ref{sec:fitres} that the spectral weight of the $\rho$
meson is significantly reduced at a temperature of $200$MeV, we conclude that the medium can no longer be thought of
as consisting of weakly interacting hadrons flying past each other.

In the high-temperature phase, the data in Table \ref{tab:chisv^2}
shows that the static susceptibility $\chi_s$ approaches unity in
units of $T^2$, as expected. The expected quasiparticles at
sufficiently high temperatures are quarks and gluons and we find a
relatively low value of the effective rms velocity. A possible
explanation is that no well-defined transport peak exists at
$T=254\,$MeV and $T=340$MeV. And if the transport peak is present, the width of the resolution function 
may be too broad to yield an accurate estimate of the rms quasiparticle velocity.

%%%%%%%%%%%%%%%%%%%%%%%%%%%%%%%%%%%%%%%%%%%%%%%%%%%%%%%%%%%%%%%%
\begin{figure}[t!]
\centering
\includegraphics[width=0.49\textwidth]{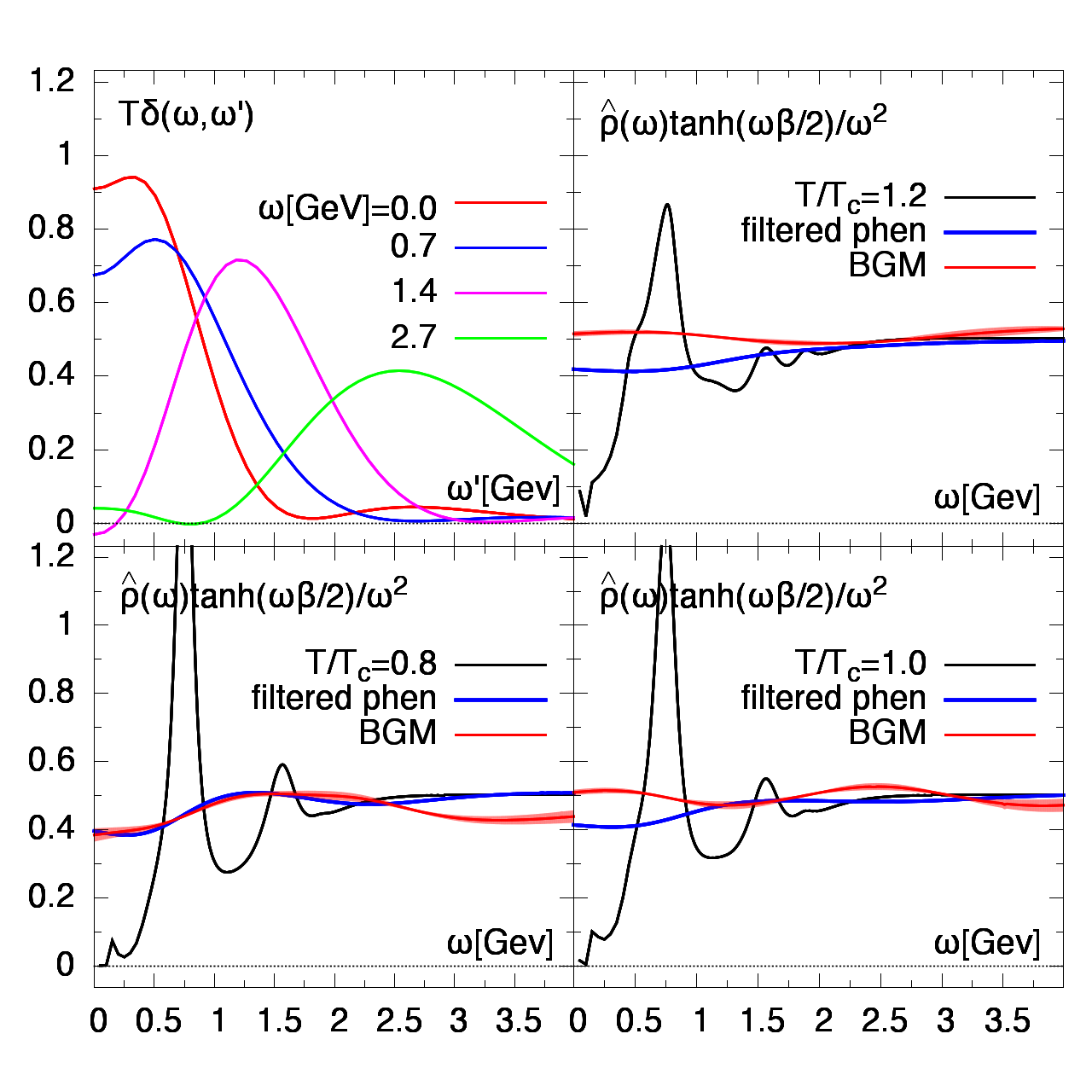}
\includegraphics[width=0.49\textwidth]{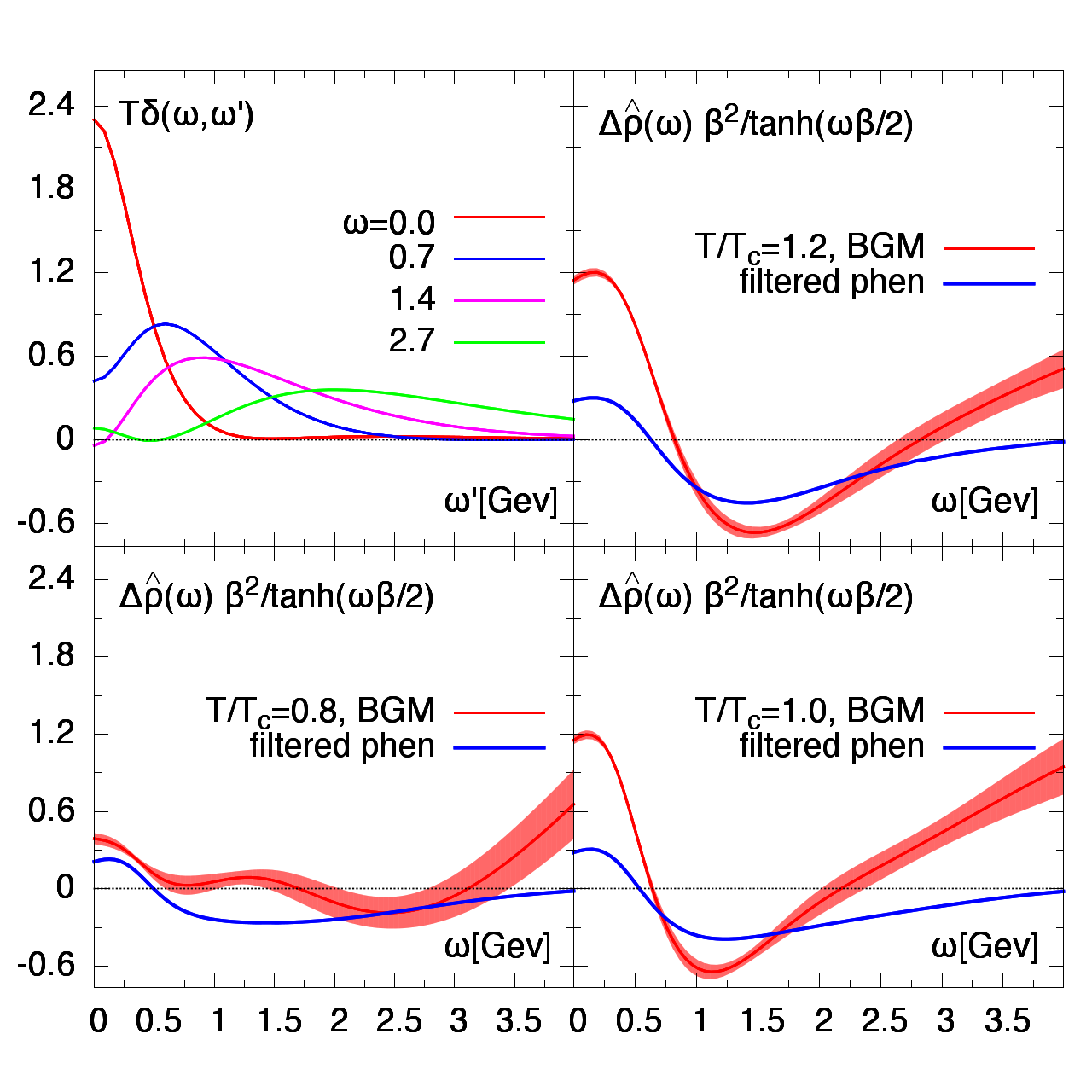}
\caption{Left: Comparison of filtered spectral functions
  $\widehat{\rho}(\omega)$ obtained by convoluting with the resolution
  functions $\delta(\omega,\omega')$ determined from the lattice,
  (red) BGM reconstruction, (blue) spectral functions from
  phenomenology \cite{Hohler:2013eba}. For comparison, the
  un-filtered result is displayed in black. The top left panel shows the resolution
  functions. Right: The same plot for the difference
  $\Delta\widehat{\rho}(\omega)$. Note, the phenomenological spectral
  functions do not contain a transport contribution, we therefore do
  not expect good agreement in the low frequency region at $T>T_c$.  }
\label{fig:phencomp}
\end{figure}

\subsection{Comparison with the thermal spectral functions of Hohler and Rapp\la{sec:pheno}}

In \cite{Hohler:2013eba} the authors used QCD and Weinberg sum rules
to study thermal, isovector vector and axial-vector spectral functions
across the phase transition. The in-medium condensates, required in
this analysis, were estimated using the hadron resonance gas model and
lattice QCD data where possible. In the following, we compare these
phenomenological spectral functions with those determined from our
lattice correlation functions. However, in such a comparison a number
of issues have to be considered, e.g. the pion masses are larger than
in nature and the calculations presented here are not in the continuum
limit. In addition, the lattice spectrum in a finite volume consists
of a discrete set of delta functions. A direct comparison between
spectral functions is therefore not meaningful. On the other hand,
comparing Euclidean correlation functions obscures the physical
interpretation, which is normally based on the local behavior of the
spectral function around a given frequency. Hence the BGM provides an
opportunity to compare phenomenological and lattice calculations via
the filtered phenomenological spectral function.  Indeed, all that is
needed for this comparison is the resolution function.
A possible recipe to compare $\widehat\rho(\omega)$ is to
filter the given spectral functions $\rho_{\rm{phen}}(\omega')$
with the resolution functions determined by the lattice data,
\be\la{eq:smear}
 \widehat{\rho}_{\rm{phen}}(\omega) = \int_0^\infty d\omega' \delta(\omega,\omega')\rho_{\rm{phen}}(\omega')~~.
\ee

To map the phenomenological curves of \cite{Hohler:2013eba} to our
results, we set $T_c=155$MeV in QCD with physical $u,d,s$ quark
masses~\cite{Borsanyi:2010bp,Bazavov:2011nk}, entailing the following
temperatures $T_{\rm phen}$ where the phenomenological spectral functions will
be evaluated,
\ba
N_\tau = 24:~~~&T/T_c \approx 0.8 &~~\Rightarrow~~ T_{\rm phen}=130\rm{MeV}~,\nonumber\\
N_\tau = 20:~~~&T/T_c \approx 1.0 &~~\Rightarrow~~ T_{\rm phen}=150\rm{MeV}~,\nonumber\\
N_\tau = 16:~~~&T/T_c \approx 1.25 &~~\Rightarrow~~ T_{\rm phen}=170\rm{MeV}~.%\\\nonumber
\ea
Furthermore we normalize phenomenological curves to the lattice
spectral functions by dividing out the charge factor $C_{\rm em}=5/9$.
In the left panel of Fig.~\ref{fig:phencomp} we show the results for
the filtered spectral functions $\widehat\rho(\omega)$ and in the
right one those for the difference $\Delta\widehat{\rho}(\omega)$. We
present our results for the BGM (red), the filtered phenomenological (blue)
spectral functions and the unfiltered phenomenological spectral
functions (black) in Fig.~\ref{fig:phencomp} (left). As a transport
contribution is not included in the phenomenological data, we do not
expect very good agreement in the low frequency region for
temperatures above $T_c$. Nevertheless, the qualitative agreement between
the phenomenological and reconstructed lattice results is good. Even
though a full quantitative study has to be postponed, 
 the presented method allows for a systematic comparison
 of BGM reconstructed lattice spectral functions
and phenomenological results in the future.

%%%%%%%%%%%%%%%%%%%%%%%%%%%%%%%%%%%%%%%%%%%%%%%%%%%%%%%%%%%%%%%%
%%%%%%%%%%%%%%%%%%%%%%%%%%%%%%%%%%%%%%%%%%%%%%%%%%%%%%%%%%%%%%%%

\section{Electrical conductivity across the phase transition}
\la{sec:implic}

In this section we discuss the impact of our results on our knowledge
of the electrical conductivity across the
deconfinement phase transition. We recall that the electrical
conductivity can be read off from the spectral functions via
Eq.~(\ref{eq:sigD}) from the intercept of $\rho(\omega)/\omega$ at
$\omega=0$. Since the BGM gives access only to the filtered spectral
functions $\widehat\rho(\omega)$, we cannot determine a model
independent result for $\sigma_{\rm el}$ at this time.  Nonetheless,
we present the results obtained from the ansatz approach in Tab.~\ref{tab:sigma}. 
We quote central values obtained with (Mod. 2c), and use (Mod. 2d) as a way
to estimate the systematic error from the choice of ansatz.

Here the errors are determined from the spectral functions for
(Mod. 2c) and (Mod. 2d). The first value denotes those from (Mod. 2c),
the second gives the lower bounds from (Mod. 2d) relative to the
central values of (Mod. 2c) and the third the corresponding upper
bound. We observe and increase of the electrical conductivity across
the phase transition.

%%%%%%%%%%%%%%%%%%%%%%%%%%%%%%%%%%%%%%%%
\begin{table}[t!]
\begin{tabular}{llcl@{~~~}}
\hline\hline\noalign{\smallskip}
$T[{\rm MeV}]$~~~    &   $T/T_c$~~~  &  $\sigma_{\rm el}/C_{\rm em} T$ &  $\Delta\sigma_{\rm el}/C_{\rm em} T$  \\
\noalign{\smallskip}\hline\noalign{\smallskip}
169   &   0.8   & 0.074 & (61)(-74)(+0) \\
203   &   1.0   & 0.160 & (6)(-29)(+40) \\
254   &   1.25   & 0.189 & (10)(-13)(+4) \\
338   &   1.67   & 0.323 & (46)(-78)(+78) \\
\noalign{\smallskip}\hline\hline
\end{tabular}
\caption{The electrical conductivity $\sigma_{\rm el}$ determined via the ansatz approach. The central value (third column) 
is obtained with (Mod.\ 2c). The uncertainties, displayed in the fourth column, are determined from the bootstrap samples.
The first value is the statistical error of (Mod. 2c).
The second and third give the range of values (relative to the central value of the third column)
obtained with (Mod. 2d).}
\label{tab:sigma}
\end{table}
%%%%%%%%%%%%%%%%%%%%%%%%%%%%%%%%%%%%%%%%

\begin{figure}[t!]
\centering
\includegraphics[width=0.49\textwidth]{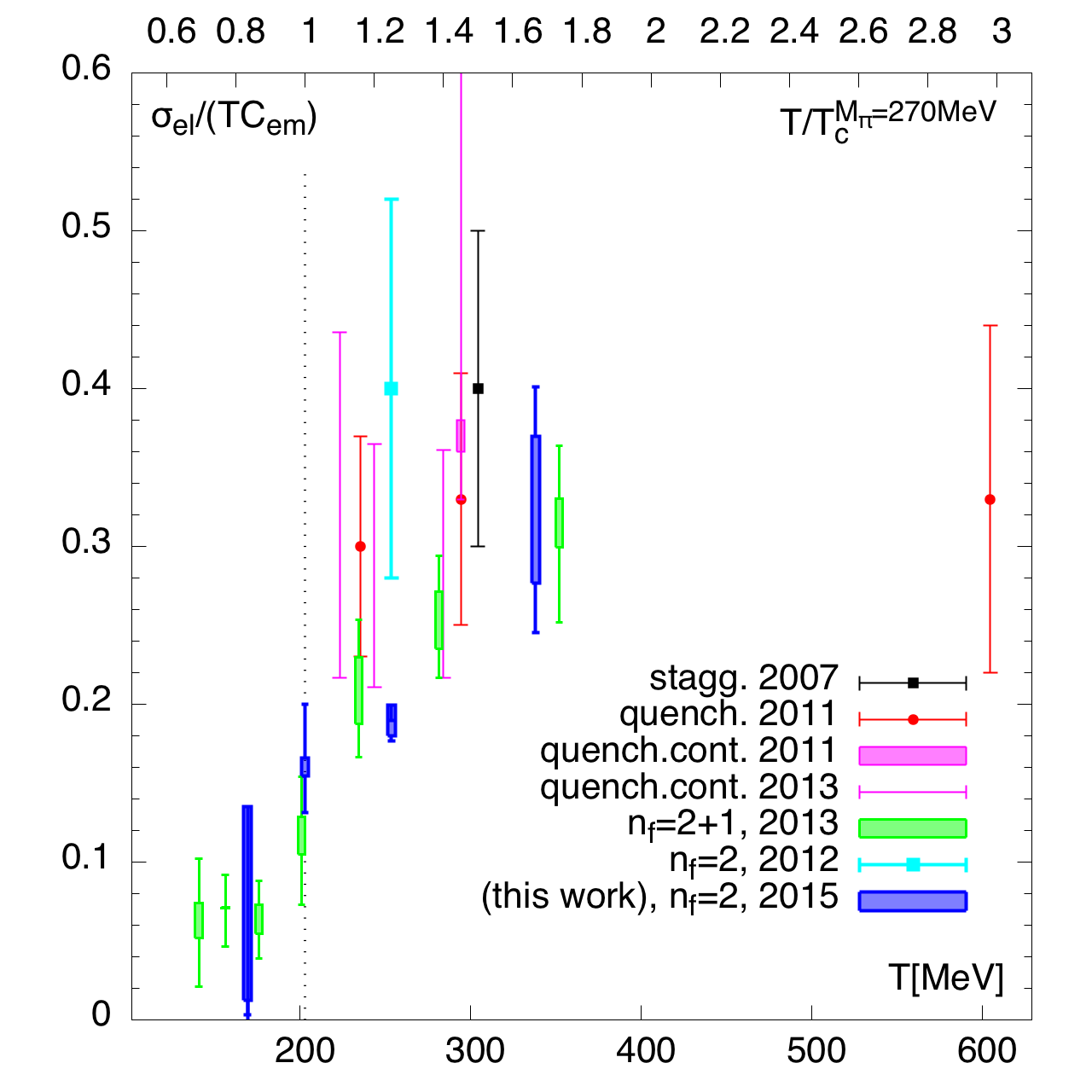}
\caption{The electrical
  conductivity read off from the fitted spectral functions. For our quoted results (blue) the boxes give the errors determined from (Mod. 2c) and the errorbars those
  from (Mod. 2d).
  The results from this work are compared to past staggered (grey)
  \cite{Aarts:2007wj}, quenched (red) \cite{Francis:2011bt}, quenched
  in the continuum limit (blue triangles and magenta bars)
  \cite{Ding:2010ga,Ding:2013qw,Ding:2014dua}, $n_f=2+1$ (green
  whiskerbars) \cite{Amato:2013naa}, as well as our $n_f=2$
  \cite{Brandt:2012jc} studies. }
\label{fig:sigma}
\end{figure}

Both the ansatz method and MEM introduce a significant degree of
systematic uncertainty, due to the ill-posed nature of the inverse
transformation. Consequently cross-checks using multiple approaches
are mandatory. In Fig.~\ref{fig:sigma} we collect and compare
different determinations of $\sigma_{\rm el}$ from recent lattice
determinations using staggered quenched (grey) \cite{Aarts:2007wj},
Wilson-Clover quenched (red) \cite{Francis:2011bt}, Wilson-Clover
quenched in the continuum limit (blue triangles and magenta
bars)\cite{Ding:2010ga,Ding:2013qw,Ding:2014dua}, $n_f=2+1$ (green
whiskerbars)\cite{Amato:2013naa}, as well as our previous $n_f=2$
(blue) Wilson-Clover studies \cite{Brandt:2012jc}. Here the staggered
and Wilson-Clover $n_f=2+1$ results use MEM for the reconstruction,
while the Wilson-Clover quenched and our $n_f=2$ studies use the
ansatz method.  Note, in our previous study \cite{Brandt:2012jc} we
computed the local-local current correlators with a factor 64 lower
statistics. Additionally the model did not permit the separation of the
contributions for the thermal and vacuum bound states. In the quenched
calculations no clear temperature dependence can be seen, while there
is a consistent drop in dynamical QCD. This might
be due to the different nature of the deconfinement phase transitions
in both theories. Comparing especially with the $n_f=2+1$ MEM \cite{Amato:2013naa} dynamical results, the
results are in good agreement. A consistent picture is emerging
for the  electrical conductivity around the critical temperature $T_c$ from lattice simulations. 

\section{Summary and outlook\la{sec:concl}}

We have presented the Euclidean correlation functions in the isovector
vector channel computed using lattice QCD at temperatures $T/T_c=0.8,\;
1.0,\; 1.25$ and $1.67$. We have analyzed the correlation functions in
terms of the associated spectral functions.  The simulations were
performed with two flavors of light quarks and included a reference
simulation at zero temperature. In this setup, the pseudocritical temperature
amounts to $T_c=203\,$MeV.
With high precision Monte-Carlo data available for all lattice
ensembles, we applied a set of sophisticated ans\"atze and exploited an exact sum rule
(\eq(\ref{eq:sr})) constraining the vacuum-subtracted spectral function to describe
the underlying spectral functions.

Fairly general arguments based on kinetic theory suggest that the
$\rho$ meson is largely unaffected by thermal effects as long as the
conserved-charge fluctuations and the spectral weight of the transport
peak are as small as predicted by the hadron resonance gas model (section
\ref{sec:kin}). 
By fitting the ansatz parameters to the lattice data, we find that the
spectral weight below the zero-temperature threshold $\omega=2m_\pi$  remains small at $T=0.8T_c$,
but becomes significant at $ T_c$.
This thermal effect is accompanied by a reduction of the spectral weight above the threshold,
particularly  around the vector meson mass
$m_\rho$. This may be interpreted as evidence of a rapid dissociation
of the vector meson as the system crosses into the deconfined phase.
At $T=T_c$ we find that the hadron resonance gas
predictions fail to describe the charge fluctuations (section \ref{sec:vsq}).
For $T>T_c$, the fitted spectral functions do not exhibit
any particular excess of spectral weight around the $\rho$ meson mass (Fig.~\ref{fig:spf}).

Using the appropriate Kubo formula, we estimated the
electrical conductivity across the deconfinement phase transition
(Fig.~\ref{fig:sigma}) and compared it with recent lattice
QCD calculations. We observe a rapid increase of the electrical
conductivity across the deconfinement phase transition in line with that observed in other dynamical calculations. A consistent picture is emerging.

In addition to providing a phenomenologically motivated form of the
spectral function which successfully describes the lattice correlation function data, we
applied the Backus-Gilbert method to the lattice data. The main advantage of
this method is that a `filtered' spectral function is constructed,
which, at any given frequency, is given as a local weighted average of
the true spectral function with a known weight function.  In this method 
the spectral function does not have to be expanded in any specific basis of functions and
the ill-posed nature of the inverse problem translates into the limited
locality in frequency that can be achieved in the relation between the
filtered and the actual spectral function. We also point out that the
filtered spectral function obtained using the Backus-Gilbert method,
does not necessarily provide a description of the lattice data with a 
small $\chi^2$, once integrated with the appropriate kernel in \eq(\ref{eq:coshK}).
In that sense, the method is complementary to applying
a fit ansatz for the spectral function, or to applying the maximum
entropy method.

Having obtained the filtered spectral functions 
at $T=0.8T_c$, we compared the low-frequency region to the predictions
of the hadron resonance gas model (section \ref{sec:vsq}) and found
agreement at the $10$--$15\%$ level. 
The agreement is somewhat improved if a pion gas is assumed with a modified
dispersion relation at low momenta~\cite{Brandt:2015sxa}.
At $T=T_c$, as already stated
above, the hadron resonance gas model fails. Secondly, we compared the filtered spectral
functions to previously published \cite{Hohler:2013eba}
phenomenological spectral functions. Once they are passed through the
same filter, they exhibit the same qualitative feature of having
excess spectral weight around $\omega=0$ and a depletion at and above
the $\rho$ meson mass, as compared to the respective vacuum spectral
functions. Quantitatively, we see differences, which could be partly
understood by the fact that the zero-temperature pion mass is $m_\pi\simeq270$MeV
in our lattice simulations, rather than the physical value of
$m_\pi=140$MeV. We think that comparing the filtered spectral functions
is a good way of testing phenomenological models, since it does not introduce
model-dependence on the lattice side and is on the other hand as local in
frequency as the quality of the lattice data allows.

Looking to the future, it would be highly desirable to perform a
continuum extrapolation of the lattice data before undertaking
spectral analyses. Second, achieving high resolution in frequency
space requires very high accuracy. We estimate that cleanly separating
the transport contribution from the $\rho$ meson spectral weight at
$T=0.8T_c$ in the Backus-Gilbert method would require reducing the
error bars by a factor four.  Finally, the finite-size effects on the
filtered spectral function should be studied in more detail
(qualitatively, the effects were discussed in~\cite{Meyer:2011gj}).

%%%%%%%%%%%%%%%%%%%%%%%%%%%%%%%%%%%%%%%%%%%%%%%%%%%%%%%%%%%%%%%%
%%%%%%%%%%%%%%%%%%%%%%%%%%%%%%%%%%%%%%%%%%%%%%%%%%%%%%%%%%%%%%%%
% \newpage
\acknowledgments{ 
  We thank Ralf Rapp and Paul M. Hohler for discussions and for providing their spectral functions.
  We are grateful for the access to the
  zero-temperature ensemble used here, made available to us through
  CLS. We acknowledge the use of computing
  time for the generation of the gauge configurations on the \emph{JUGENE} and \emph{JUQUEEN}
  computers of the Gauss Centre for Supercomputing located at
  Forschungszentrum J\"ulich, Germany. Part of the configurations and all correlation functions were computed on the
  dedicated QCD platforms ``\emph{Wilson}'' at the Institute for Nuclear
  Physics, University of Mainz, and ``\emph{Clover}'' at the Helmholtz-Institut Mainz. This work was supported by the
  \emph{Center for Computational Sciences in Mainz} as part of the
  Rhineland-Palatinate Research Initiative and by the DFG grant ME
  3622/2-1 \emph{Static and dynamic properties of QCD at finite
    temperature}.
}

%%%%%%%%%%%%%%%%%%%%%%%%%%%%%%%%%%%%%%%%%%%%%%%%%%%%%%%%%%%%%%%%
%%%%%%%%%%%%%%%%%%%%%%%%%%%%%%%%%%%%%%%%%%%%%%%%%%%%%%%%%%%%%%%%
%\clearpage
\appendix

\section{Correlation function data}
\la{app:data}

Below we list the results of $G_{ii}(\tau,T)$ correlators up to the
midpoint, the corresponding covariance matrices of these data sets
will be provided online \cite{arxiv}. Due to the use of the
conserved-local current $G_{00}(\tau,T)$ is constant to very high accuracy
and can be inferred from $\chi_s/T^2$ in Tab.~\ref{tab:chisv^2}.

\begin{table}[h!]
\centering
 % Give a unique label
% For LaTeX tables use
\begin{tabular}{lllc|clllc|clllc|clll  }
\hline\hline\noalign{\smallskip}
\multicolumn{18}{c}{$64^3\times 128$, $T=32(6)$MeV}\\
\noalign{\smallskip}\hline\noalign{\smallskip}
$\tau$ & $G_{ii}(\tau,T)$ & $\delta G_{ii}(\tau,T)$ &&& $\tau$ & $G_{ii}(\tau,T)$ & $\delta G_{ii}(\tau,T)$ &&& $\tau$ & $G_{ii}(\tau,T)$ & $\delta G_{ii}(\tau,T)$ &&& $\tau$ & $G_{ii}(\tau,T)$ & $\delta G_{ii}(\tau,T)$\\
\noalign{\smallskip}\hline\noalign{\smallskip}
0 & -1.6350e-01 &2.998e-04 &&& 16 & 3.6539e-05 &4.461e-07 &&& 32 & 1.2809e-06 &1.412e-07 &&& 48 & 1.1428e-08 &4.920e-08 \\
1 & 5.2587e-02 &1.092e-04 &&& 17 & 2.9049e-05 &4.042e-07 &&& 33 & 1.0345e-06 &1.344e-07 &&& 49 & 1.8294e-08 &4.908e-08 \\
2 & 1.6643e-02 &3.171e-05 &&& 18 & 2.3136e-05 &3.782e-07 &&& 34 & 8.2386e-07 &1.251e-07 &&& 50 & 8.3935e-09 &4.811e-08 \\
3 & 5.9285e-03 &1.084e-05 &&& 19 & 1.8501e-05 &3.531e-07 &&& 35 & 6.4831e-07 &1.177e-07 &&& 51 & 5.4551e-09 &4.589e-08 \\
4 & 2.6076e-03 &5.721e-06 &&& 20 & 1.4874e-05 &3.278e-07 &&& 36 & 5.1108e-07 &1.095e-07 &&& 52 & 7.8290e-09 &4.511e-08 \\
5 & 1.3495e-03 &3.332e-06 &&& 21 & 1.1931e-05 &2.989e-07 &&& 37 & 3.9113e-07 &1.028e-07 &&& 53 & 9.9049e-09 &4.458e-08 \\
6 & 7.8590e-04 &2.203e-06 &&& 22 & 9.6562e-06 &2.745e-07 &&& 38 & 2.8859e-07 &9.439e-08 &&& 54 & 2.5509e-08 &4.406e-08 \\
7 & 4.9835e-04 &1.561e-06 &&& 23 & 7.8557e-06 &2.564e-07 &&& 39 & 2.2771e-07 &8.799e-08 &&& 55 & 3.9952e-08 &4.336e-08 \\
8 & 3.3602e-04 &1.192e-06 &&& 24 & 6.3920e-06 &2.467e-07 &&& 40 & 1.8372e-07 &8.136e-08 &&& 56 & 5.7284e-08 &4.183e-08 \\
9 & 2.3710e-04 &9.685e-07 &&& 25 & 5.2335e-06 &2.335e-07 &&& 41 & 1.4408e-07 &7.643e-08 &&& 57 & 6.9352e-08 &4.223e-08 \\
10 & 1.7282e-04 &8.212e-07 &&& 26 & 4.3140e-06 &2.132e-07 &&& 42 & 1.0570e-07 &7.222e-08 &&& 58 & 7.7771e-08 &4.250e-08 \\
11 & 1.2917e-04 &7.202e-07 &&& 27 & 3.5265e-06 &2.007e-07 &&& 43 & 7.4499e-08 &6.713e-08 &&& 59 & 8.4635e-08 &4.303e-08 \\
12 & 9.8227e-05 &6.312e-07 &&& 28 & 2.8993e-06 &1.880e-07 &&& 44 & 5.1063e-08 &6.236e-08 &&& 60 & 9.2662e-08 &4.395e-08 \\
13 & 7.5773e-05 &5.802e-07 &&& 29 & 2.3682e-06 &1.725e-07 &&& 45 & 4.1019e-08 &5.864e-08 &&& 61 & 1.0136e-07 &4.602e-08 \\
14 & 5.9029e-05 &5.481e-07 &&& 30 & 1.9094e-06 &1.581e-07 &&& 46 & 2.3223e-08 &5.499e-08 &&& 62 & 9.7743e-08 &4.693e-08 \\
15 & 4.6229e-05 &4.982e-07 &&& 31 & 1.5624e-06 &1.507e-07 &&& 47 & 1.4130e-08 &5.221e-08 &&& 63 & 9.7347e-08 &4.820e-08 \\
     &  					&					&&&		  &					 & 				 &&&      &					  &				  &&& 64 & 9.7577e-08 &4.932e-08 \\
\noalign{\smallskip}\hline\hline
% \noalign{\smallskip}\hline
\end{tabular}
\caption{{The vacuum vector correlators with ensemble parameters $64^3\times 128$, $T=32(6)$MeV, $N_{src}=16$, $m_\pi\simeq 270$MeV and $m_\pi L=4.2$.}}
\label{tab:vacdat}      
\end{table}

\begin{table}[h!]
\centering
 % Give a unique label
% For LaTeX tables use
\begin{tabular}{lllc|clllc|clllc|clll  }
\hline\hline\noalign{\smallskip}
\multicolumn{4}{c|}{$64^3\times 24$, $T=169(3)$MeV }&\multicolumn{5}{c|}{$64^3\times 20$, $T=203(4)$MeV}&\multicolumn{5}{c|}{$64^3\times 16$, $T=254(4)$MeV}&\multicolumn{4}{c}{$64^3\times 12$, $T=338(5)$MeV }\\
\noalign{\smallskip}\hline\noalign{\smallskip}
$\tau$ & $G_{ii}(\tau,T)$ & $\delta G_{ii}(\tau,T)$ &&& $\tau$ & $G_{ii}(\tau,T)$ & $\delta G_{ii}(\tau,T)$ &&& $\tau$ & $G_{ii}(\tau,T)$ & $\delta G_{ii}(\tau,T)$ &&& $\tau$ & $G_{ii}(\tau,T)$ & $\delta G_{ii}(\tau,T)$\\
\noalign{\smallskip}\hline\noalign{\smallskip}
0 & -1.6399e-01 &1.163e-04 &&& 0 & -1.6430e-01 &1.014e-04 &&& 0 & -1.6463e-01 &1.001e-04 &&& 0 & -1.6477e-01 &1.084e-04 \\
1 & 5.2705e-02 &4.293e-05 &&& 1 & 5.2719e-02 &3.575e-05 &&& 1 & 5.2776e-02 &3.694e-05 &&& 1 & 5.2798e-02 &4.143e-05 \\
2 & 1.6667e-02 &1.277e-05 &&& 2 & 1.6692e-02 &1.311e-05 &&& 2 & 1.6768e-02 &1.045e-05 &&& 2 & 1.6963e-02 &1.160e-05 \\
3 & 5.9540e-03 &5.149e-06 &&& 3 & 5.9975e-03 &4.781e-06 &&& 3 & 6.0897e-03 &4.161e-06 &&& 3 & 6.3814e-03 &4.744e-06 \\
4 & 2.6307e-03 &2.711e-06 &&& 4 & 2.6814e-03 &2.401e-06 &&& 4 & 2.7971e-03 &2.282e-06 &&& 4 & 3.1940e-03 &2.763e-06 \\
5 & 1.3775e-03 &1.725e-06 &&& 5 & 1.4382e-03 &1.515e-06 &&& 5 & 1.5797e-03 &1.528e-06 &&& 5 & 2.1229e-03 &2.166e-06 \\
6 & 8.1961e-04 &1.262e-06 &&& 6 & 8.9125e-04 &1.140e-06 &&& 6 & 1.0665e-03 &1.197e-06 &&& 6 & 1.8551e-03 &2.063e-06 \\
7 & 5.3822e-04 &1.076e-06 &&& 7 & 6.2263e-04 &9.765e-07 &&& 7 & 8.4697e-04 &1.106e-06 &&&  &  &\\
8 & 3.8362e-04 &9.826e-07 &&& 8 & 4.8443e-04 &8.882e-07 &&& 8 & 7.8483e-04 &1.188e-06 &&&  &  &\\
9 & 2.9444e-04 &9.473e-07 &&& 9 & 4.1733e-04 &8.890e-07 &&&  &  &&&&  &  &\\
10 & 2.4309e-04 &9.308e-07 &&& 10 & 3.9737e-04 &8.967e-07 &&&  &  &&&&  &  &\\
11 & 2.1627e-04 &9.299e-07 &&&  &  &&&&  &  &&&&  &  &\\
12 & 2.0807e-04 &9.329e-07 &&&  &  &&&&  &  &&&&  &  &\\
\noalign{\smallskip}\hline\hline
% \noalign{\smallskip}\hline
\end{tabular}
\caption{{The thermal vector correlators on the lattice sizes $64^3\times 24, 20, 16$ and 12 using the fixed scale approach.}}
\label{tab:thermdat}      
\end{table}
%%%%%%%%%%%%%%%%%%%%%%%%%%%%%%%%%%%%%%%%%%%%%%%%%%%%%%%%%%%%%%%%
%%%%%%%%%%%%%%%%%%%%%%%%%%%%%%%%%%%%%%%%%%%%%%%%%%%%%%%%%%%%%%%%

%\clearpage
%\vspace{1cm}
% \newpage
%%%%%%%%%%%%%%%%%%%%%%%%%%%%%%%%%%%%%%%%%%%%%%%%%%%%%%%%%%%%%%%%
\bibliography{dileptons.bib}
%%%%%%%%%%%%%%%%%%%%%%%%%%%%%%%%%%%%%%%%%%%%%%%%%%%%%%%%%%%%%%%%

\end{document}